\documentclass[
preprint, 
 amsmath,amssymb,
 aps, physrev,
]{revtex4-2}

\usepackage{graphicx}% Include figure files
\usepackage{dcolumn}% Align table columns on decimal point
\usepackage{bm}% bold math
\usepackage{hyperref}% add hypertext capabilities
\usepackage{adjustbox}
\usepackage{subcaption}
\usepackage{float}
\usepackage{booktabs}
\usepackage{tabularx}
\newcolumntype{L}{>{\raggedright\arraybackslash}X}

\begin{document}

\preprint{APS/123-QED}

\title{\textbf{Co-evolution of cooperation and resource allocation in the advantageous environment-based spatial multi-game using adaptive control} 
}

\author{Chengbin Sun}
\email{Contact author: sunchengbin94@163.com}
\affiliation{
  Department of Economics and Management, Dalian University of Technology, Dalian 116024, China
}
\affiliation{
  Institute for Biocomputation and Physics of Complex Systems (BIFI), University of Zaragoza, Zaragoza 50018, Spain.
}

\author{Alfonso de Miguel-Arribas}
\affiliation{
 Zaragoza Logistics Center (ZLC), Zaragoza 50018, Spain.
}

\author{Chaoqian Wang}
\affiliation{Department of Computational and Data Sciences, George Mason University, Fairfax, VA 22030, USA.}

\author{Haoxiang Xia}
\email{Contact author: hxxia@dlut.edu.cn}
\affiliation{
  Department of Economics and Management, Dalian University of Technology, Dalian 116024, China.
}

\author{Yamir Moreno}
\affiliation{
 Institute for Biocomputation and Physics of Complex Systems (BIFI), University of Zaragoza, Zaragoza 50018, Spain
}
\affiliation{
 Department of Theoretical Physics, University of Zaragoza, Zaragoza 50018, Spain
}
\affiliation{
 Centai Institute, Turin, Italy.
}

\date{\today}

\begin{abstract}
In real-life complex systems, individuals often encounter multiple social dilemmas that cannot be effectively captured using a single-game model. Furthermore, the environment and limited resources both play a crucial role in shaping individuals’ decision-making behaviors. In this study, we employ an adaptive control mechanism by which agents may benefit from their environment, thus redefining their individual fitness. Under this setting, a detailed examination of the co-evolution of individual strategies and resource allocation is carried. Through extensive simulations, we find that the advantageous environment mechanism not only significantly increases the proportion of cooperators in the system but also influences the resource distribution among individuals. Additionally, limited resources reinforce cooperative behaviors within the system while shaping the evolutionary dynamics and strategic interactions across different dilemmas. Once the system reaches equilibrium, resource distribution becomes highly imbalanced. To promote fairer resource allocation, we introduce a minimum resource guarantee mechanism. Our results show that this mechanism not only reduces disparities in resource distribution across the entire system and among individuals in different dilemmas but also significantly enhances cooperative behavior in higher resource intervals. Finally, to assess the robustness of our model, we further examine the influence of the advantageous environment on system-wide cooperation in small-world and random graph network models.
\end{abstract}

\maketitle

\section{Introduction}
Cooperation is a widespread phenomenon in nature, observed across various species, from the collective hunting strategies of lions and wolves to the intricate social behaviors of human societies. However, the prevalence of cooperation appears paradoxical when viewed through the lens of natural selection, which favors individual fitness~\cite{darwin1859origin, packer1991molecular, fehr2018normative}. This suggests that the inherent self-interest of individuals poses a fundamental challenge to the evolution of cooperation. Understanding the emergence and persistence of cooperative behavior among self-interested agents remains a central question in diverse fields, including biology~\cite{burnham2005biological}, social sciences~\cite{fowler2010cooperative}, and behavioral sciences~\cite{gintis2004towards}. Evolutionary game theory provides a powerful framework for addressing this issue~\cite{nowak2004evolutionary, pennisi2005did, perez2020complex}, with various game-theoretic models widely employed, such as the Prisoner’s Dilemma~\cite{mcnamara2004variation}, the Snowdrift Game~\cite{hauert2004spatial}, the Stag Hunt~\cite{wang2013evolving}, and the Public Goods Game~\cite{wang2021public}.

The Prisoner's Dilemma Game (PDG) and the Snowdrift Game (SDG) are two fundamental two-player, two-strategy game models, each representing a distinct social dilemma. In both games, two players simultaneously choose between two strategies: cooperation (C) or defection (D). If both players cooperate, they each receive a reward payoff $R$; if both defect, they receive a punishment payoff $P$. When one player cooperates while the other defects, the cooperator incurs the sucker’s payoff $S$, whereas the defector obtains the temptation payoff $T$. In the PDG, the parameters satisfy $T > R > P > S$ along with $2R > T + S$, meaning that defection is always the dominant strategy, leading to mutual defection as the equilibrium outcome. In contrast, in the SDG, the parameters satisfy $T > R > S > P$, implying that a player benefits most by choosing the strategy opposite to their opponent's choice. In recent years, research has increasingly focused on multi-game models that integrate both PDG and SDG, yielding important insights. For instance, Huang {\it et al.}~\cite{huang2019evolution} introduced desire-driven update rules in multi-game settings and analyzed their impact on the evolution of cooperation. Roy {\it et al.}~\cite{roy2022eco} proposed a multi-game theoretical model that elucidates mechanisms for maintaining biodiversity. Furthermore, Liu {\it et al.}~\cite{liu2022effect} investigated how perceived competition and learning costs influence cooperation in multi-game systems, demonstrating that these factors can significantly enhance the prevalence of cooperative behavior.

 In the past few decades, numerous mechanisms have been proposed to explain the emergence and persistence of cooperative behavior. Notably, Nowak summarized five key rules~\cite{nowak2006five}: kin selection, direct reciprocity, indirect reciprocity, group selection, and network reciprocity. Among these, network reciprocity has been shown to significantly promote cooperation within a system, attracting substantial research interest~\cite{floria2009social,semmann2012conditional,li2023social}. In structured populations, individuals interact only with their immediate neighbors, and cooperators can resist the invasion of defectors by forming cooperative clusters~\cite{nowak1992evolutionary}. The evolution of cooperation has been extensively studied across various network structures, including small-world networks~\cite{deng2010memory,lin2020evolutionary}, scale-free networks~\cite{santos2005scale,kleineberg2017metric}, interdependent networks~\cite{wang2013interdependent,wu2023evolution}, signed networks~\cite{song2022evolutionary}, and higher-order networks~\cite{xu2024reinforcement,civilini2021evolutionary,alvarez2021evolutionary}. Additionally, various mechanisms such as aspiration~\cite{wang2019evolutionary}, memory~\cite{lu2018role}, reputation~\cite{wang2012inferring}, punishment~\cite{henrich2006cooperation}, and migration~\cite{zhang2022migration} have been explored to better understand the emergence and stability of cooperation. Furthermore, the influence of external factors on individual decision-making in different systems has gained significant attention, leading to important findings~\cite{guo2017environment,jin2017incorporating,tilman2020evolutionary}. However, existing studies have yet to examine the impact of an advantageous environment on cooperative behavior under multiple dilemmas. In real-world scenarios, when individuals face multiple dilemmas, they aim to minimize losses and mitigate risks. Rational individuals will only consider environmental factors if the environmental payoff exceeds their personal payoff. In this study, we refer to the social neighborhood of an individual in the system as the social environment, or environment for short, and define an environment as advantageous when an individual's environmental payoff is higher than their own payoff. 
 
 In multiple dilemmas, not only does the social environment influence individual decision-making, but an individual's resources also play a crucial role in shaping their strategic behavior. Here, resources refer to the material necessities required for the survival of organisms and individuals in both nature and human society, such as food, water, habitat, and money. These resources are essential for survival and have played a fundamental role in biological evolution. The primary objective of living organisms is to acquire as many resources as possible to ensure survival and reproduction. However, limited resources often lead to competition, particularly among individuals or groups. In competitive environments, cooperative behavior can emerge spontaneously as a means to maintain collective benefits and defend against external threats. Thus, resource dynamics is inherently intertwined with strategic interactions: resource acquisition depends on game outcomes, and the amount of resources an individual possesses influences their strategic choices. Moreover, competition for resources can create substantial imbalances, making unfair distribution a pressing issue of concern in modern society. To address the challenge of equitable resource allocation, scholars across various disciplines have conducted extensive research. For example, Loumiotis {\it et al.}~\cite{loumiotis2014dynamic} investigated the allocation of backhaul network resources in base stations using evolutionary game theory, modeling interactions between users and base stations. Similarly, Semasinghe {\it et al.}~\cite{semasinghe2014evolutionary} proposed a distributed resource allocation scheme based on evolutionary game theory (EGT) for small cells within a macro-cellular network. Additionally, Hazarika {\it et al.}~\cite{hazarika2022drl} introduced a priority-sensitive task offloading and resource allocation scheme for Internet of Vehicles networks.

In this paper, we propose a multi-game model incorporating an advantageous social environment mechanism using an adaptive control method, as well as resource dynamics under limited resources . Specifically, a coupling parameter $\mu$ to the agent's social environment is introduced into individuals’ payoffs, redefining their fitness when environmental payoffs exceed individual payoffs. Using this model, we analyze the co-evolution of strategies and the distribution of resources. First, we examine how limited resources and the advantageous social environment influence the fraction of cooperation under various system scenarios. Second, we explore the impact of the initial distribution of game-based subpopulations, and the role of the sucker’s payoff in determining the fraction of cooperation. Next, we analyze in detail the effects of the coupling parameter $\mu$, the sucker’s payoff $\sigma$, and the temptation to defect $b$ on the cooperation fraction. Furthermore, to gain deeper insight into the evolutionary dynamics in the various games, we investigate the relationship between strategy proportions, evolutionary trajectories, and average fitness under several $\mu$ and $\sigma$. Additionally, we examine how resource distribution behaves across the overall system, within the different game-based subpopulation, and how impacts strategy adoption. We also assess the impact of varying $\mu$ and $\sigma$ on resource dynamics at both the system and individual levels. Finally, to evaluate the robustness of our proposed advantageous environment mechanism, we study the evolutionary dynamics of the advantageous environmental mechanism within cooperative systems structured on small-world and random graph networks.  

This paper is organized as follows. Section~\ref{sec:model} introduces the proposed multi-game model, with adaptive control based on the advantageous environment mechanism, and limited resource dynamics. Section~\ref{sec:results} presents the main results and discussions. Finally, we summarize our findings and conclusions in Section~\ref{sec:discussion}.

\section{Model}
\label{sec:model}

Our model consists of a set of players located on a regular square lattice of size $N=L\times L$ with periodic boundary conditions and a von Neumann neighborhood (i.e., degree $k=4$ for all players). Each site in the lattice represents a player, and connections between sites denote interactive relationships. 

At the initial stage, players are randomly assigned a game to play, and thus they are divided into two subpopulations. We denote by $\Upsilon_{PDG}$ the set of players playing PDG, comprising a fraction $1-\omega$ of the whole system, and by $\Upsilon_{SDG}$ the set of players following SDG, with a fraction $\omega$. Following previous research~\cite{li2019effect}, we use the so-called weak PDG and SDG, with the following payoff matrices:

\begin{equation}
{Q}=\begin{pmatrix}R&S\\T&P\end{pmatrix},~T=b\text{ ($1<b\leq2$)},~R=1,~P=0,~S=\pm\sigma\text{ ($0\leq\sigma\leq1$)}.
\end{equation}

In the PDG, the parameters are given by $R=1$, $P=0$, $T=b$ ($1\leq b\leq2$), and $S=-\sigma$ ($0<\sigma\leq1$). The SDG parameters are rescaled such that $S=+\sigma$ ($0<\sigma\leq1$), while the other values remain unchanged. These can be represented as:

\begin{equation}
{Q_1}=\begin{pmatrix}1&-\sigma\\b&0\end{pmatrix},~{Q_2}=\begin{pmatrix}1&+\sigma\\b&0\end{pmatrix}.
\end{equation}

According to previous research~\cite{tanimoto2007relationship}, the relationship between parameters is given by:

\begin{equation}
D_{g}=T-R,~D_{r}=P-S,~D_{g}^{^{\prime}}=\frac{D_{g}}{R-P},~D_{r}^{^{\prime}}=\frac{D_{r}}{R-P}.
\end{equation}

From (1), (2), and (3), it follows that $D_{g}=D_{g}^{^{\prime}}=b-1>0$ and $D_{r}=D_{r}^{^{\prime}}=\pm\sigma$. When $\sigma=0$, all players participate in a weak PDG, marking the boundary between PDG and SDG. When $\sigma\neq0$, both PDG and SDG coexist in the system~\cite{wang2015universal}.

At the initial stage, each agent is allocated one unit of resources. Each player adopts either a cooperative ($S_{x}$) or defective ($S_{y}$) strategy with equal probability, expressed as:

\begin{equation}
S_x=(1,0)^\top,~S_y=(0,1)^\top.
\end{equation}

\begin{figure}
    \centering
    \includegraphics[width=0.75\textwidth]{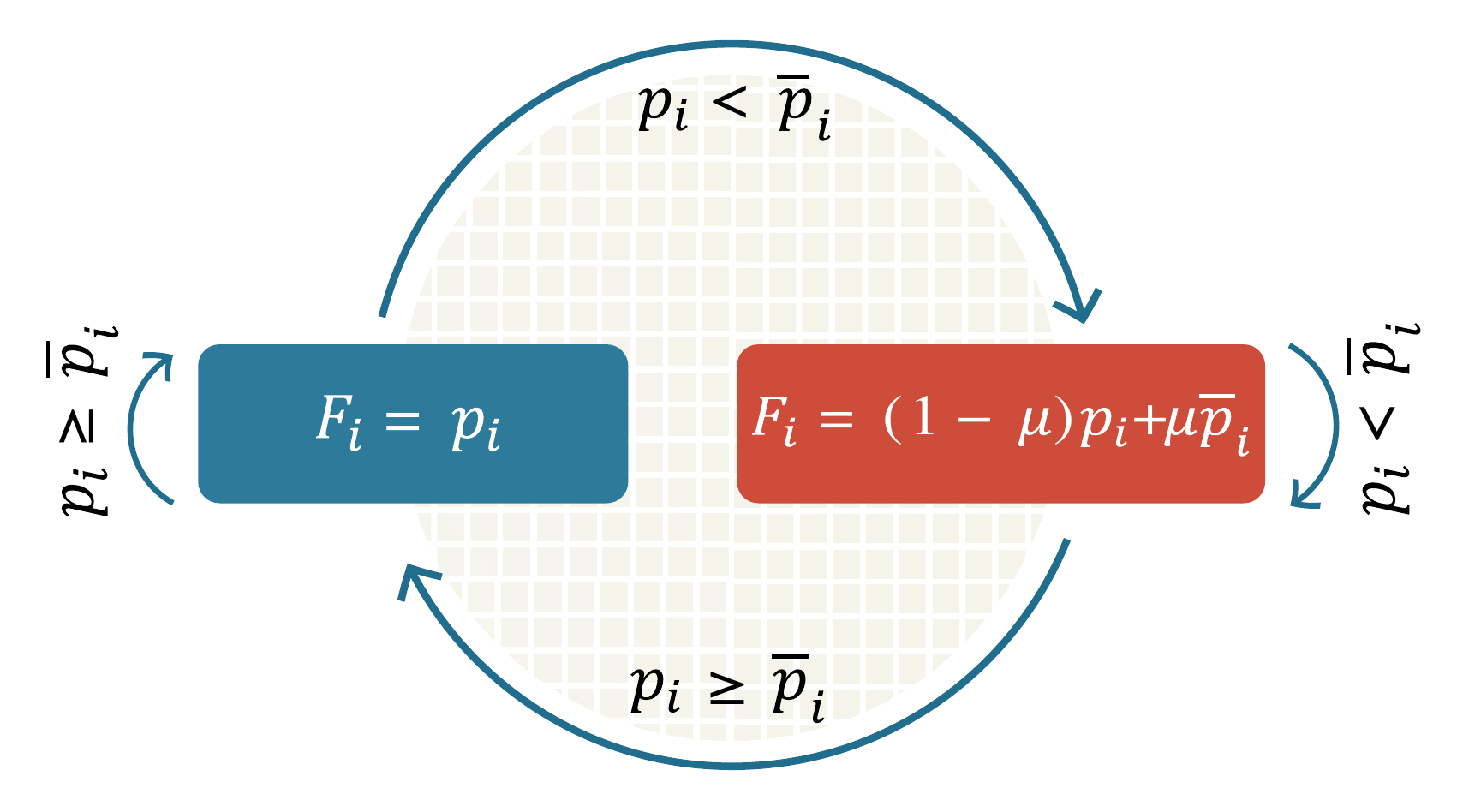}
    \caption{The adaptive control mechanism of an individual's fitness in the presence of an advantageous environment. The arrows represent the adaptive control process, with the labels indicating the relationship between $p_{i}$ and $\bar{p}_{i}$ required for each control step. When $\mu=0$, all individuals play traditional SDG and PDG, without considering the influence of the advantageous environment. When $\mu\neq0$ and $p_{i}<\bar{p}_{i}$, the advantageous environment is introduced into the multi-game model. Furthermore, when $\mu=1$ and $p_{i}<\bar{p}_{i}$, the player's fitness is entirely determined by the advantageous environment.}
    \label{figwtheat1}
\end{figure}

The system evolves through a standard discrete-time asynchronous Monte Carlo simulation, consisting of the following steps:  
\begin{enumerate}
    \item Game stage. Following typical Glauber dynamics, a player $i$ is randomly selected and obtains a payoff $p_{i}$ by engaging in games with its four nearest neighbors, regardless of their game type. This is given by:
    \begin{equation}\label{eqpiit}
    p_i=\sum_{j\in\Omega(i)}S_iQ_{T_i}S_j,
    \end{equation}
    where $\Omega(i)$ denotes the set of neighboring individuals of $i$, whose cardinality is just the agent's degree $|\Omega(i)|=k_i=4$, and $Q_{T_{i}}$ is the payoff matrix corresponding to the individual $i$'s game type.

    \item Social environment assessment. In both natural and social systems, the external factors, i.e. the environment, influences individual payoffs and decision-making. Following previous studies~\cite{guo2017environment,jin2017incorporating}, the social environment is defined as:
    \begin{equation}
    \bar{p}_i=\frac{\sum_{j\in\Omega(i)}^{k_i}p_j}{k_i},
    \end{equation}
    where $k_{i}=4$ represents the degree of player $i$. A favorable environment has a positive impact on nature and society, enhancing individual survival, whereas an unfavorable environment has the opposite effect~\cite{mahaputra2022factors}. If $p_{i}<\bar{p}_{i}$, then $\bar{p}_{i}$ is termed an advantageous environment; otherwise, it is a disadvantageous environment. Note that to obtain $p_j$ at this step, we also make each player $j$ in turn to play with their respective neighbors.
    
    \item Adaptive control and advantageous environment mechanism. To mitigate the adverse effects of disadvantageous environments on individual behavior when facing social dilemmas, we introduce the advantageous environment into the individual’s payoff. Individuals are endorsed with a fitness function on which the adaptive control mechanism operates (See Fig.~\ref{figwtheat1}). The fitness function for the focal agent $i$ is defined as:
    \begin{equation}
    F_i =
    \begin{cases} 
        p_i, & \text{if } p_i \geq \bar{p}_i, \\ 
        (1-\mu)p_i + \mu\bar{p}_i, & \text{if } p_i < \bar{p}_i.
    \end{cases}
    \end{equation}
    If the player performs better than their social peers, their fitness comes exclusively from their own contribution. Otherwise, the adaptive control mechanism operates. Then, the parameter $\mu$ ($0\leq\mu\leq1$) represents the coupling of the individual to its social neighborhood or environment. If $\mu\to 0$, again fitness comes mainly from the individual's own performance, but as $\mu\to 1$, the importance of the advantageous environment to the player's fitness increases.  In the limit $\mu=1$, the player's fitness depends exclusively on their neighborhood performance (fitness obtained through advantageous environment).
    \item Strategy-resource exchange and update. A neighboring player $j$ of the focal player $i$ is randomly selected, and their fitness $F_{j}$ is determined as given by the previous equation. Player $i$ then attempts to propagate its strategy to $j$ at a rate given by the modified Fermi function:
    \begin{equation}
    W_{(i\to j)}=\frac{r_i}{1+\exp((F_j-F_i)/k)},
    \end{equation}
    where $k=0.1$ represents the noise effect in the strategy update process, also referred to as the selection strength~\cite{yu2016system}. The term $r_{i}$ denotes the resource quantity of player $i$. Regarding resource evolution, each player initially possesses one unit of resources. When player $i$ successfully propagates its strategy to neighbor $j$, it gains a fraction $\varepsilon$ of $j$'s resources, while player $j$ loses an equal amount. Following previous studies~\cite{luo2017coevolving,luo2021environment}, we set $\varepsilon=10\%$ in our simulations unless otherwise specified. In case of a successful strategy and resource exchange, they are consequently updated for the involved agents and the dynamics proceeds again to the first step in the list (Game stage).
\end{enumerate}

All Monte Carlo simulation results presented in this study are based on networks of sizes ranging from $\mathrm{200\times200}$ to $\mathrm{600\times600}$ players. The final proportion of cooperators is measured over the last $10^{4}$ full steps of a total of $10^{5}$ steps. To ensure statistical accuracy, the reported results are averaged over $50$ independent realizations for each parameter configuration.

\section{Results}
\label{sec:results} 

\begin{figure}
    \centering
    \includegraphics[width=0.75\textwidth]{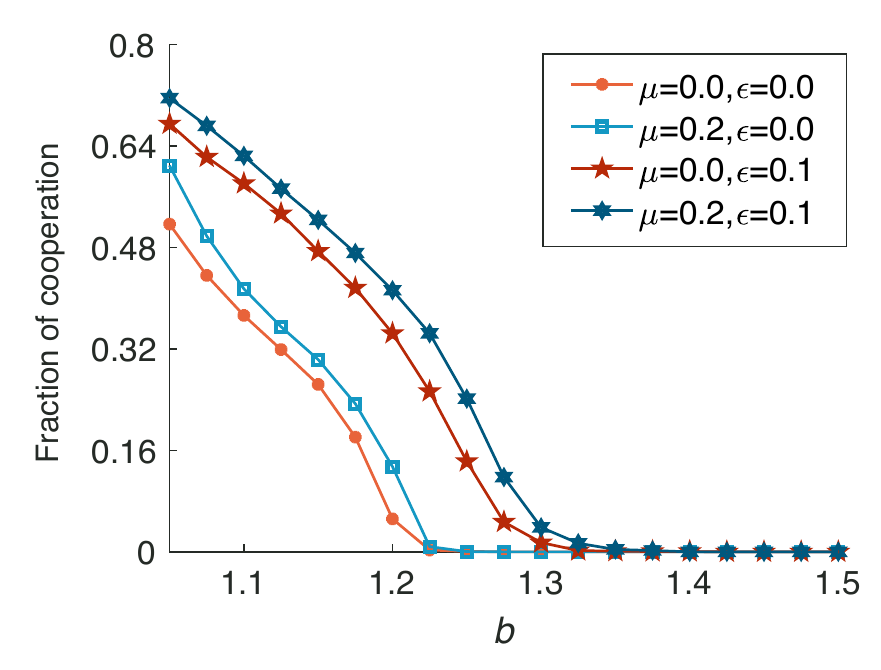}
    \caption{Cooperation fraction as a function of the temptation to defect $b$ in the multi-game system for individuals with and without an advantageous environment and resource transfer dynamics. The following scenarios are compared: (i) Players without an advantageous environment ($\mu=0$) and without resources ($\epsilon=0)$; (ii) players with an advantageous environment ($\mu=0.2$) but without resource transfer ($\epsilon=0$); (iii) players without an advantageous environment ($\mu=0$) but with resources transfer ($\epsilon=0.1$); (iv) players with both an advantageous environment ($\mu=0.2$) and resources ($\epsilon=0.1$). Other fixed relevant parameters in the simulation are $\omega=0.5$, $\sigma=0.2$.}
    \label{figwtheat2}
\end{figure}

We begin with a comparative experiment to assess the effect that the presence/absence of the combination of an advantageous social environment and resource transfer have on the macroscopic fraction of cooperation, as shown in Fig.~\ref{figwtheat2}. In the presence of resource dynamics ($\epsilon=0.1$), the existence of an advantageous environment enhances cooperation relative to the case where there is not such an advantage ($\mu=0$). The same occurs in the absence of any resource transfer $(\epsilon=0)$, where the case with $\mu=0.2$ shows a somewhat higher cooperation than that of $\mu=0$. In any case, an increasing temptation parameter $b$ leads to a decrease in cooperation. As cooperation levels collapse to zero, the critical value of $b_c$, separating these phases coincides, respectively, for the scenario with and without resource transfer. We find that $b_c\approx 1.225$ for the cases $\mu=0.2$ and $\mu=0$ with $\epsilon=0$, and $b_c\approx 1.325$ for the cases $\mu=0.2$ and $\mu=0$ with $\epsilon=0.1$. Compared with previous studies~\cite{deng2018multi}, our findings demonstrate that an advantageous social environment, induced by the adaptive control mechanism, and the presence of resource transfer significantly enhance cooperation within the system. Moreover, both factors effectively increase the threshold $b_c$ at which cooperation vanishes, making cooperation more resilient to increasing temptation levels.  

Having established the positive effect of both, a nonzero coupling to the advantageous social environment ($\mu\neq 0$) through adaptive control and the resource transfer dynamics, we now investigate how the fraction of players participating in each game (determined by $\omega$) and the sucker's payoff $\sigma$ influence cooperation levels in the system.  

\begin{figure}
    \centering
    \includegraphics[width=1.0\textwidth]{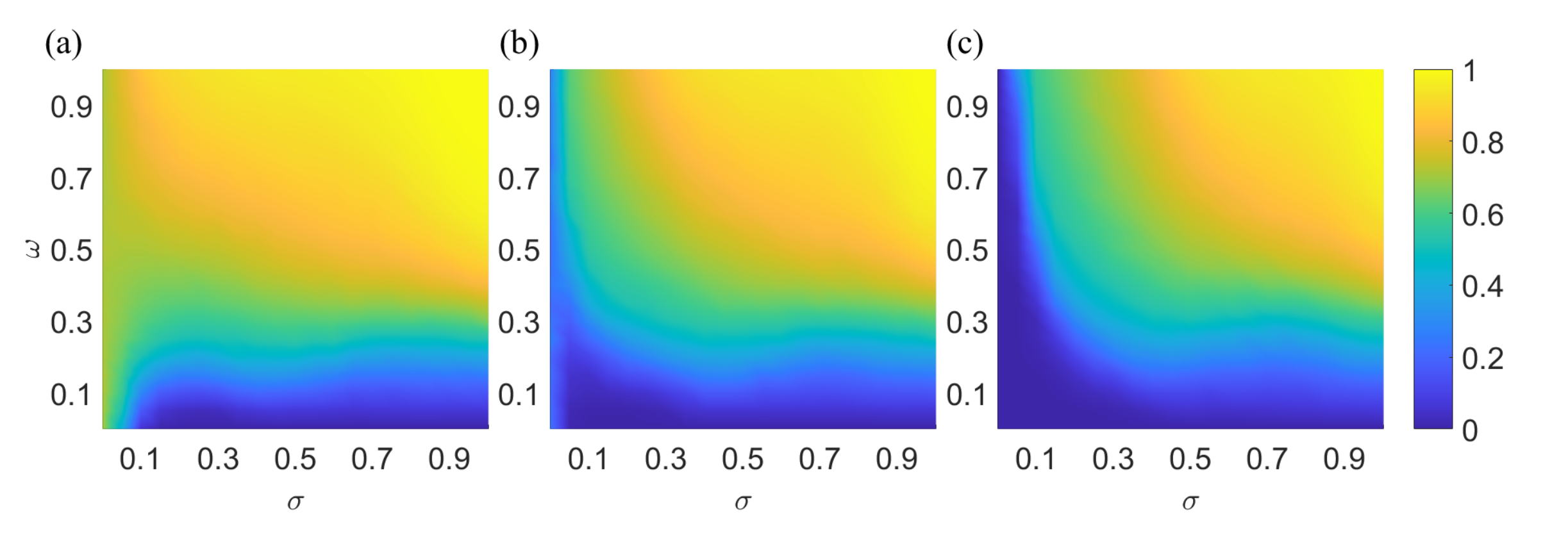}
    \caption{Fraction of cooperation heatmaps in $(\sigma,\omega)$ control parameter space for selected values of temptation parameter $b$. Panel (a) $b=1.05$. Panel (b) $b=1.10$. Panel (c) $b=1.15$. For all the scenarios it is set $\mu=0.2$.}
\label{figwtheat3}
\end{figure}

In Fig.~\ref{figwtheat3} a collection of heatmaps showing the fraction of cooperators in control parameter space $(\sigma,\omega)$ is depicted, for selected values of $b$: (a) $b=1.05$ , (b) $b=1.10$, and (c) $b=1.15$. We can found that as expected, increasing $b$ negatively impacts cooperation levels. Indeed, this change makes cooperation collapse in the region of low $\sigma$, for any $\omega$, as can be observed from Fig.~\ref{figwtheat3}(a) to Fig.~\ref{figwtheat3}(c). Consistently, with low to very low $\omega$, cooperation is also zero for practically any $\sigma$ value (except for very low $\sigma$ and $b=1.05$). Contrarily, across any section of $b$, cooperation levels stay at around $0.5$ or higher for the medium-low to high $\omega$ and the entire range of $\sigma$, except very low values as $b$ is increased. As the fraction of players playing SDG increases (high $\sigma$) and the payoff obtained in $C$-$D$ interactions of the SDG grows positively ($+\omega$), cooperation is promoted and ends up being massively adopted across the system. Compared to traditional single game models~\cite{luo2017coevolving}, our multi-game framework combined with the adaptive control mechanism for advantageous social environment creates an expanded parameter space that fosters cooperation. In the subsequent analyses, we will assume the number of players belonging to $\Upsilon_{SDG}$ and $\Upsilon_{PDG}$ stays fixed and equal to $\omega=0.5$, thus half of the system plays SDG and the other half plays PDG.

\begin{figure}
    \centering
    \includegraphics[width=1.0\textwidth]{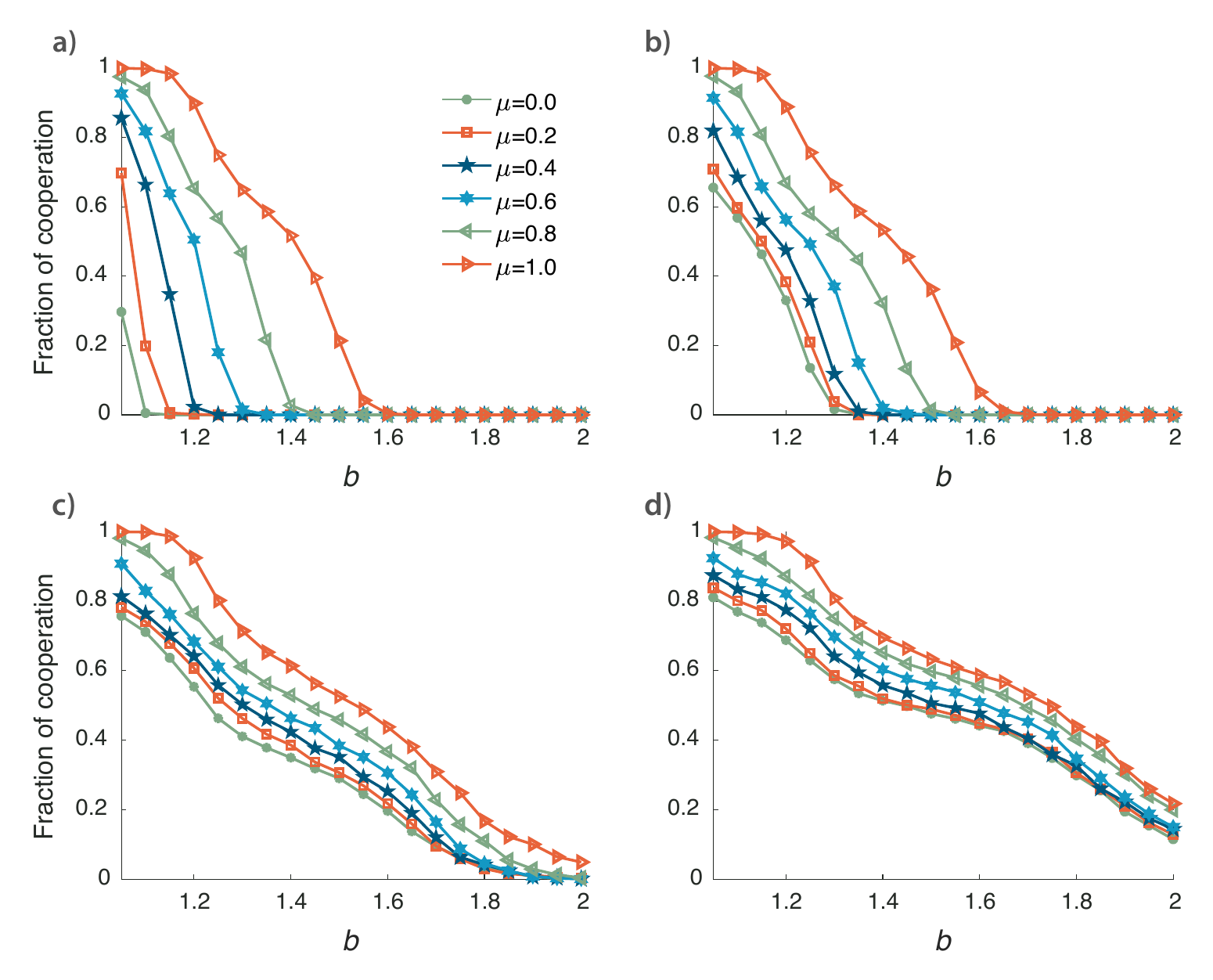}
    \caption{Fraction of cooperation as a function of $b$ for a variety of $\mu$ values, at selected sucker's payoff $\sigma$: (a) $\sigma=0$, (b) $\sigma=0.2$, (c) $\sigma=0.5$, (d) $\sigma=0.8$.}
    \label{figwtheat4}
\end{figure}

Now, we delve into the impact of the social coupling parameter $\mu$ and the payoff $\sigma$ on the macroscopic fraction of cooperation. Fig.~\ref{figwtheat4} reveals the behavior of the cooperation fraction as a function of $b$ for a variety of $\mu$ values and selected $\sigma$: (a) $\sigma=0$, (b) $\sigma=0.2$, (c) $\sigma=0.5$, (d) $\sigma=0.8$. In the limit case of $\sigma=0$, the totality of players in the system are actually playing the weak PDG. Compared to previous research~\cite{luo2017coevolving}, we can observe the positive effects of the parameter $\mu$. Continuing with Fig.~\ref{figwtheat4}(a),  we observe that when individual behavior dominates ($\mu=0$), the fraction of cooperation is very low ($\approx 0.3$) even for very small $b$ and quickly decays to zero for increasing $b$. However, on the other extreme, $\mu=1.0$, cooperation still dominates at $b$ values as high as $b=1.4$. Moving to higher $\sigma$, Fig.~\ref{figwtheat4}(b) to (d), we see how this change also promotes cooperation. Cooperation fractions are systematically higher as $b$ increases, subsequently, the value of $b_c$ at which cooperation vanishes is increased. In the case of $\sigma=0.8$, cooperation can be remarkably sustained for high $b$. In sum, in a system where half of the players play PDG and the other half play SDG, the social coupling parameter $\mu$ promotes cooperation, and cooperation levels are further fostered as $\sigma$ increases. Even though increasing $\sigma$ affects negatively cooperators in $C$-$D$ interactions in PDG, it has the opposite effect for them in $C$-$D$ interactions within SDG.

\begin{figure}
    \centering
    \includegraphics[width=1.0\textwidth]{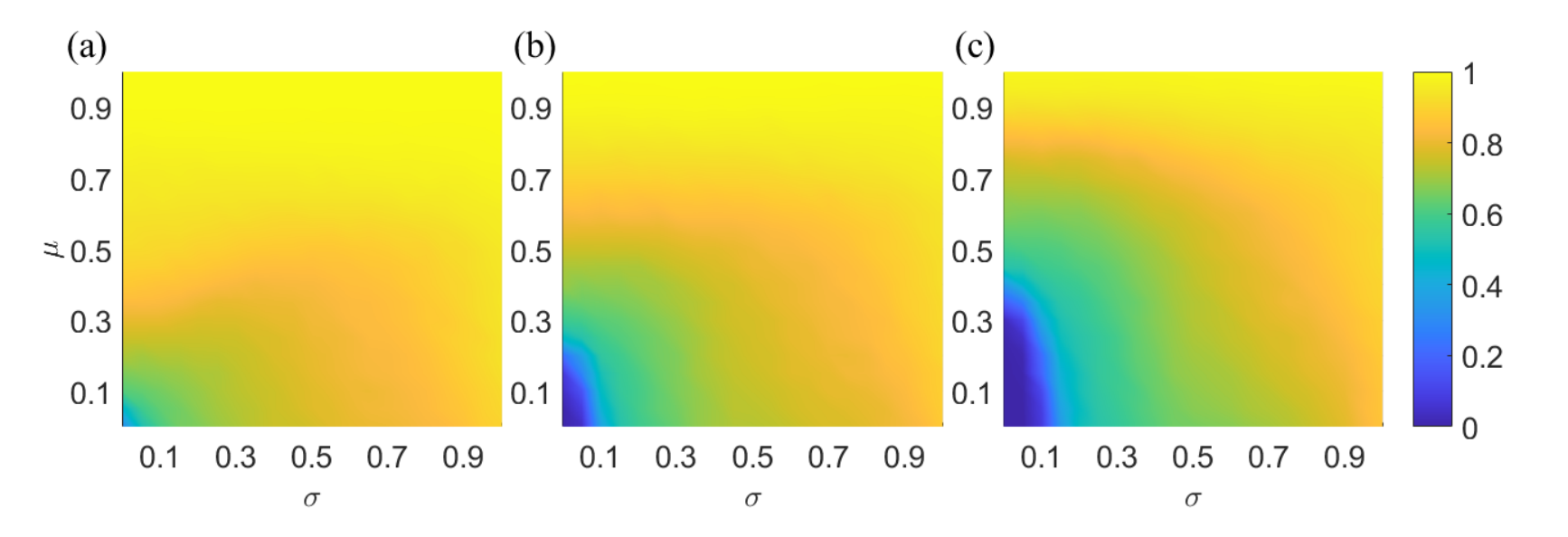}
    \caption{Fraction of cooperation heatmaps in $(\sigma,\mu)$ control parameter space for selected values of temptation parameter $b$. Panel (a) $b=1.05$. Panel (b) $b=1.10$. Panel (c) $b=1.15$.}
    \label{figwtheat5}
\end{figure}

To further characterize the influence of $\mu$ and $\sigma$ on cooperation levels, in Fig.~\ref{figwtheat5}, we perform an exhaustive exploration in $(\sigma,\mu)$ control parameter space for selected values of $b$: (a) $b=1.05$, (b) $b=1.10$, and (c) $b=1.15$, respectively. As expected, only as $b$ increases and for low $\sigma$ and $\mu$, cooperation levels vanishes. Conversely, the majority of $(\sigma,\mu)$ promotes high levels of cooperation. Cooperation is totally adopted for high $\mu$, across any value of $\sigma$. Compared to previous research~\cite{luo2021environment}, our multi-game framework can sustain higher cooperation at higher $b$ values.

\begin{figure}
    \centering
    \includegraphics[width=1.0\textwidth]{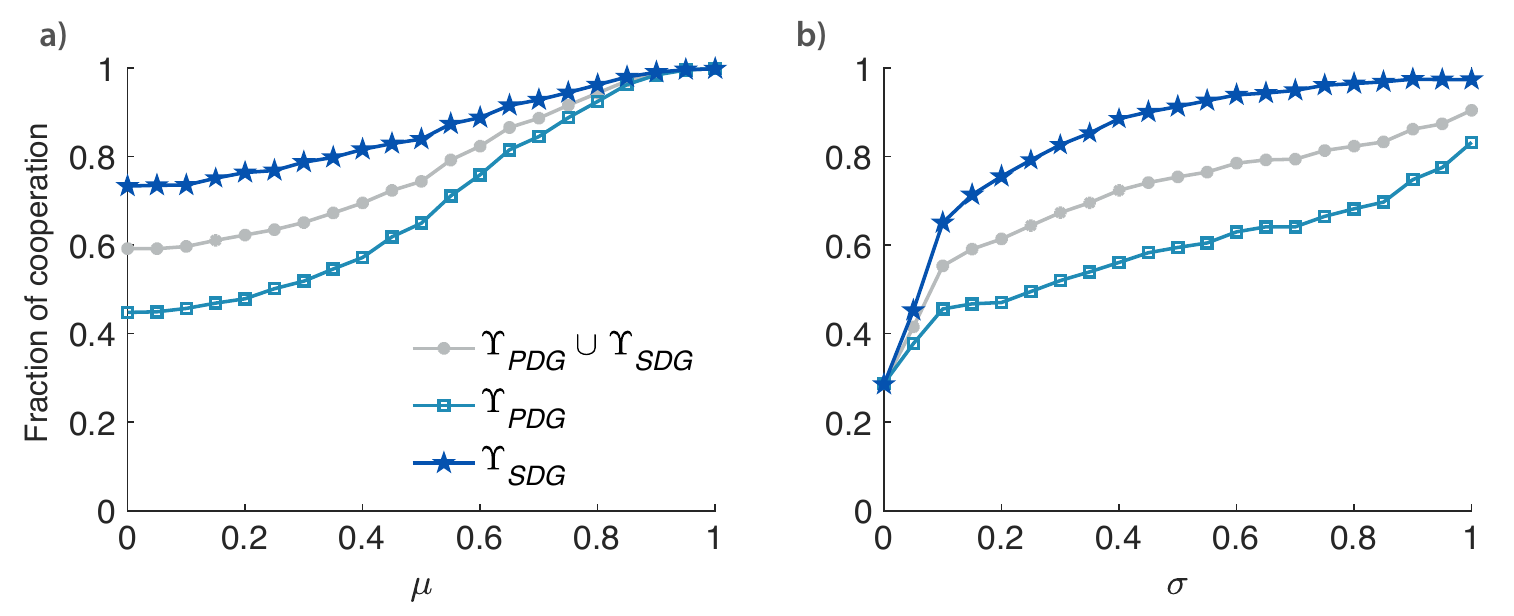}
    \caption{Fraction of cooperation as a function of $\mu$ and $\sigma$ with (a) $\sigma=0.2$ and (b) $\mu=0.2$, for the entire system, $\Upsilon_{PDG}\cup\Upsilon_{SDG}$, the population playing PDG, $\Upsilon_{PDG}$, and the population playing SDG, $\Upsilon_{SDG}$, respectively. The rest of parameter is set to be $b=1.10$.}
    \label{figwtheat6}
\end{figure}

Now, in Fig.~\ref{figwtheat6}, we delve into how cooperation levels are shared within the sets of players playing respectively PDG and SDG. For reference, we also represent the cooperation fraction in the entire system. Fig.~\ref{figwtheat6}(a) depicts this observable as a function of $\mu$, for fixed $\sigma=0.2$. Consistently, cooperation is higher in SDG relative to PDG, along the whole range of $\mu$. In the medium to low range of $\mu$, the gap is wider and, but with increasing $\mu$, the gap is narrowed as cooperation is adopted in the entire system. We can observe how the effect of the social coupling parameter is to unify the adoption of cooperation in both sets of players. When exploring the cooperation fraction with respect to $\sigma$ in Fig.~\ref{figwtheat6}(b), the same qualitative behavior is observed, now with the fraction of cooperation as a function of $\sigma$, and fixed $\mu=0.2$. When $\sigma=0$, since there is no distinction between PDG and SDG, the fraction of cooperation is the same in both sets of players, as $\sigma$ increases, the differences in games yield a wide gap between both types of players, only slightly narrowed again at the other extreme.

\begin{figure}
    \centering
    \includegraphics[width=1.0\textwidth]{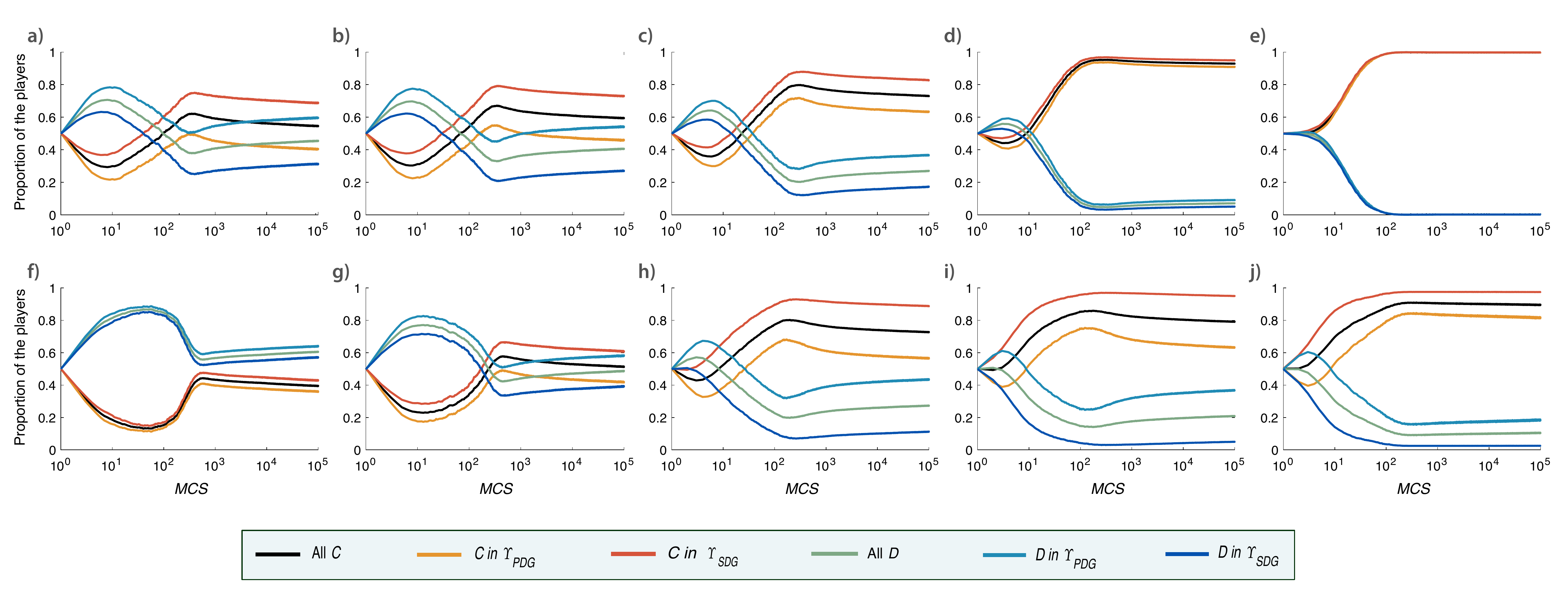}
    \caption{Time evolution in Monte Carlo Steps (MCS) of the fraction of cooperators and defectors in the entire system, and within sets $\Upsilon_{PDG}$ and $\Upsilon_{SDG}$, in the top row for (a) $\mu=0$, (b) $\mu=0.2$, (c) $\mu=0.5$, (d) $\mu=0.8$, and (e) $\mu=1.0$, all under fixed $b=1.10$ and $\sigma=0.2$. And in the bottom row, for (f) $\sigma=0.05$, (g) $\sigma=0.1$, (h) $\sigma=0.5$, (i) $\sigma=0.8$, and (j) $\sigma=1.0$, with fixed $b=1.0$ and $\mu=0.2$.}
    \label{figwtheat7}
\end{figure}

\begin{figure}
    \centering
    \includegraphics[width=1.0\textwidth]{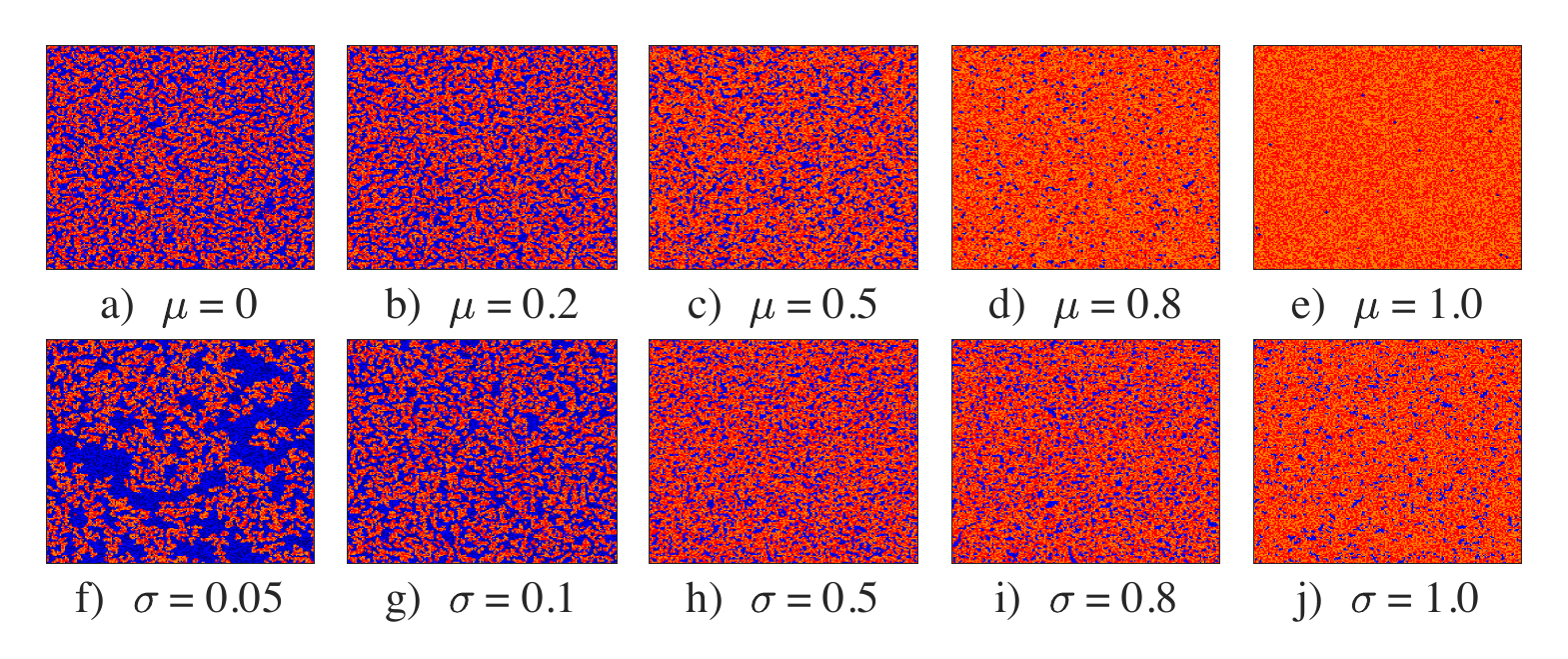}
    \caption{Strategy configuration snapshots at equilibrium (MCS $=10^5$) on the square lattice. Red and orange tiles represent cooperators playing PDG and SDG, respectively, whereas blue and dark blue tiles represent defectors respectively playing PDG and SDG. Top row depicts cases of fixed $b=1.10$ and $\sigma=0.2$, for (a) $\mu=0$, (b) $\mu=0.2$, (c) $\mu=0.5$, (d) $\mu=0.8$, and (e) $\mu=1.0$. Bottom row represents cases of fixed $b=1.10$ and $\mu=0.2$, for (f) $\sigma=0.05$, (g) $\sigma=0.1$, (h) $\sigma=0.5$, (i) $\sigma=0.8$, (j) $\sigma=1.0$.}
\label{figwtheat8}
\end{figure}

To delve into the evolutionary dynamics of the system and its different subpopulations, we analyze the evolution of cooperators and defectors across the entire system, as well as within $\Upsilon_{PDG}$ and $\Upsilon_{SDG}$, as a function of the evolutionary time, measured in Monte Carlo Steps {\it MCS}, for different values of the coupling parameter $\mu$ and the sucker’s payoff $\sigma$. The results are shown in Fig.~\ref{figwtheat7}.  

Fig.~\ref{figwtheat7}(a) --  Fig.~\ref{figwtheat7}(e) illustrate the time evolution of strategy fractions in the entire system, $\Upsilon_{PDG}$, and $\Upsilon_{SDG}$ for different values of $\mu$: $0$, $0.2$, $0.5$, $0.8$, and $1.0$. It can be observed that when $\mu=0$ and $\mu=0.2$, the fraction of cooperators (both system-wide and within $\Upsilon_{PDG}$ and $\Upsilon_{SDG}$) initially decreases, reaching a minimum at approximately $10$ {\it MCS}. As the evolutionary process unfolds, the fraction of cooperators gradually increases, peaking at around $500$ {\it MCS}, after which it stabilizes. Logically, the proportion of defectors follows an opposite trend, initially increasing and later declining as cooperation stabilizes. This is the result of a random initial distribution of strategies and agents playing each game. In this early stage, defectors, whether in PDG or SDG, exploit neighboring cooperators, obtaining higher payoffs that leads to their strategy being more easily propagated. However, as the system evolves, cooperators begin to cluster together, forming cooperative domains that resist the invasion of defectors. These clusters progressively expand, stabilizing cooperation within the system. Throughout the entire evolutionary process, the fraction of cooperators remains higher in $\Upsilon_{SDG}$ than in $\Upsilon_{PDG}$, indicating that cooperation is more sustainable within the SDG population.  

Moreover, as $\mu$ increases, the fraction of cooperators steadily grows, while the proportion of defectors declines. Additionally, the time required for the fraction of cooperators to reach its minimum and recover to its maximum decreases, suggesting that a higher $\mu$ accelerates the stabilization of cooperation. Furthermore, the gap between the proportion of cooperators in $\Upsilon_{PDG}$ and $\Upsilon_{SDG}$ narrows as $\mu$ increases, indicating a convergence in cooperation levels between the two subpopulations. The equilibrium configuration of cooperators and defectors in $\Upsilon_{PDG}$ and $\Upsilon_{SDG}$ at equilibrium is shown in Fig.~\ref{figwtheat8}(a) -- Fig.~\ref{figwtheat8}(e). These results demonstrate that as $\mu$ increases, cooperative clusters (represented by red and orange tiles) continue to expand. Notably, when $\mu=1.0$, cooperative clusters nearly span the entire system, leading to a near-complete dominance of cooperation. From this analysis, it is evident that the social coupling parameter $\mu$, induced by the adaptive control mechanism,  not only promotes cooperation within $\Upsilon_{PDG}$ and $\Upsilon_{SDG}$, but also facilitates the formation of cooperative clusters. Moreover, increasing $\mu$ reduces the initial disadvantage faced by cooperators, shortens the time required for cooperation to reach its maximum level, and accelerates the stabilization of cooperation within the system.

Now, Fig.~\ref{figwtheat7}(f) -- Fig.~\ref{figwtheat7}(j) depict the fraction of cooperators and defectors across the entire system, as well as within $\Upsilon_{PDG}$ and $\Upsilon_{SDG}$, as a function of {\it MCS} for different values of the sucker’s payoff $\sigma$. From these results, it is evident that as $\sigma$ increases, the proportion of cooperators steadily rises, while the fraction of defectors declines throughout the evolutionary process. Moreover, the fraction of cooperators in $\Upsilon_{SDG}$ consistently exceeds that in both the overall system and $\Upsilon_{PDG}$. The spatial arrangement of strategies in $\Upsilon_{PDG}$ and $\Upsilon_{SDG}$ at equilibrium for different values of $\sigma$ ($\sigma=0.05, 0.1, 0.5, 0.8, 1.0$) is shown in Fig.~\ref{figwtheat8}(f) -- Fig.~\ref{figwtheat8}(j). These results indicate that when $\sigma$ is small, cooperators remain scattered throughout the system, making them more vulnerable to defector exploitation. However, as $\sigma$ increases, cooperators progressively dominate the system, forming increasingly large cooperative clusters. Thus, we conclude that higher values of $\sigma$ promote cooperation within the multi-game framework by facilitating the formation of large cooperative clusters, which in turn enhance resistance against defectors.

\begin{table}[ht]
\scriptsize
\caption{Average fitness of cooperators and defectors in the entire system ($F_{C}$ and $F_{D}$), within $\Upsilon_{PDG}$ ($F_{PDG,C}$ and $F_{PDG,D}$), and within $\Upsilon_{SDG}$ ($F_{SDG,C}$ and $F_{SDG,D}$) for different values of {\it MCS}, considering $\mu=0$, $\mu=0.2$, $\mu=0.5$, $\mu=0.8$, and $\mu=1.0$. The main control parameters are set to $b=1.10$ and $\sigma=0.2$.}
\label{tablemcsmu}
\begin{tabularx}{\linewidth}{@{}LLLLLLLLLLLLL@{}}
\toprule
$\mu$ & {\it MCS}    & 2      & 5      & 50     & 100    & 500    & 1000   & 5000   & 10000  & 50000  & 100000        \\ 
\midrule
$\mu=0$         & {$F_{C}$} &1.7511	&2.1627 &2.6481 &2.698	&3.0587	&3.0419 &3.0016 &2.9919	&2.9648	&2.9509   \\
                & {$F_{SDG,C}$}     &2.0450	&2.4719 &2.9838 &3.1044	&3.3154	&3.2772 &3.2361 &3.2216	&3.2060	&3.1927   \\
                & {$F_{PDG,C}$}     &1.4186	&1.6855 &2.1104 &2.2648	&2.6618	&2.6613 &2.6117 &2.6046	&2.5539	&2.5364   \\
                & {$F_{D}$} &1.4186	&1.6855	&2.1104	&2.2648	&2.6618	&2.6613	&2.6117	&2.6046	&2.5539	&2.5364   \\
                & {$F_{SDG,D}$}     &1.4907	&0.9736	&1.0926	&1.3104	&1.5006	&1.4088	&1.3616	&1.3445	&1.3332	&1.3246   \\
                & {$F_{PDG,D}$}     &1.3820	&0.9516	&1.1284	&1.4484	&1.7590	&1.6414	&1.5667	&1.5360	&1.5019	&1.4817   \\   
$\mu=0.2$       & {$F_{C}$} &1.8491	&2.2522	&2.8089	&2.9600	&3.2430	&3.2239	&3.1888	&3.1729	&3.1531	&3.1461   \\
                & {$F_{SDG,C}$}     &2.1000	&2.5194	&3.0995	&3.2347	&3.4422	&3.4090	&3.3750	&3.3643	&3.374	&3.3412   \\
                & {$F_{PDG,C}$}     &1.5649	&1.8393	&2.3702	&2.5528	&2.9472	&2.9387	&2.8960	&2.8706	&2.8447	&2.8352   \\
                & {$F_{D}$} &1.5165	&1.0533	&1.2640	&1.5257	&1.8552	&1.7326	&1.6665	&1.6419	&1.6157	&1.6017   \\
                & {$F_{SDG,D}$}     &1.5630	&1.0640	&1.2216	&1.4349	&1.6845	&1.5720	&1.5337	&1.5244	&1.5109	&1.4961   \\
                & {$F_{PDG,D}$}     &1.4737	&1.0441	&1.2942	&1.5830	&1.9334	&1.8067	&1.7300	&1.6991	&1.6681	&1.6547   \\
$\mu=0.5$       & {$F_{C}$} &2.0182	&2.4220	&3.1853	&3.4002	&3.5865	&3.5654	&3.5351	&3.5273	&3.5087	&3.5050   \\
                & {$F_{SDG,C}$}     &2.2000	&2.6244	&3.3586	&3.5511	&3.6877	&3.6623	&3.6329	&3.6237	&3.6093	&3.6052   \\
                & {$F_{PDG,C}$}     &1.8155	&2.1459	&2.9669	&3.2152	&3.4588	&3.4410	&3.4081	&3.4014	&3.3769	&3.3737   \\
                & {$F_{D}$} &1.6780	&1.3291	&1.9347	&2.2781	&2.4199	&2.2962	&2.1894	&2.1679	&2.1278	&2.1111   \\
                & {$F_{SDG,D}$}     &1.7090	&1.3485	&1.9023	&2.1672	&2.3170	&2.1893	&2.1066	&2.0931	&2.0676	&2.0569   \\
                & {$F_{PDG,D}$}     &1.6493	&1.3129	&1.9573	&2.3444	&2.4624	&2.3412	&2.2263	&2.2016	&2.1559	&2.1367   \\
$\mu=0.8$       & {$F_{C}$} &2.2324	&2.6919	&3.7211	&3.8908	&3.9380	&3.9293	&3.9174	&3.9133	&3.9093	&3.9077   \\
                & {$F_{SDG,C}$}     &2.3440	&2.8275	&3.7645	&3.9131	&3.9518	&3.9430	&3.9315	&3.9279	&3.9237	&3.9223   \\
                & {$F_{PDG,C}$}     &2.1118	&2.5332	&3.6741	&3.8674	&3.9235	&3.9151	&3.9026	&3.8979	&3.8942	&3.8923   \\
                & {$F_{D}$} &1.8944	&1.8001	&2.8460	&3.2802	&3.2223	&3.0515	&2.9139	&2.8860	&2.8326	&2.8155   \\
                & {$F_{SDG,D}$}     &1.9056	&1.8211	&2.8125	&3.2515	&3.1976	&3.0550	&2.9092	&2.8863	&2.8483	&2.8263   \\
                & {$F_{PDG,D}$}     &1.8840	&1.7816	&2.8692	&3.2971	&3.2351	&3.0497	&2.9165	&2.8859	&2.8240	&2.8096   \\ 
$\mu=1.0$       & {$F_{C}$} &2.4347	&2.9686	&3.9542	&3.9936	&3.9987	&3.9987	&3.9987	&3.9987	&3.9988	&3.9987   \\
                & {$F_{SDG,C}$}     &2.4977	&3.0466	&3.9572	&3.9934	&3.9987	&3.9988	&3.9988	&3.9988	&3.9988	&3.9988   \\
                & {$F_{PDG,C}$}     &2.3701	&2.8859	&3.9510	&3.9937	&3.9987	&3.9986	&3.9987	&3.9987	&3.9987	&3.9987   \\
                & {$F_{D}$} &2.0621	&2.2156	&3.5585	&3.8278	&3.2481	&3.2420	&3.2359	&3.2332	&3.2432	&3.2450   \\
                & {$F_{SDG,D}$}     &2.0691	&2.2166	&3.5542	&3.8473	&3.2265	&3.2448	&3.2378	&3.2364	&3.2278	&3.2251   \\
                & {$F_{PDG,D}$}     &2.0553	&2.2147	&3.5620	&3.8126	&3.2703	&3.2388	&3.2339	&3.2295	&3.2600	&3.2341   \\ 
\bottomrule
\end{tabularx}
\end{table}

To gain a deeper understanding of how the coupling parameter $\mu$ influences the evolution of strategies in the different games, we analyze the average fitness of cooperators and defectors across the entire system, as well as within $\Upsilon_{PDG}$ and $\Upsilon_{SDG}$, as a function of {\it MCS} for different values of $\mu$. The results are presented in Table~\ref{tablemcsmu}. It is evident that regardless of the value of $\mu$, the average fitness of cooperators at each {\it MCS} is consistently higher than that of defectors. Notably, the average fitness of cooperators in $\Upsilon_{PDG}$ is the lowest among all groups. Additionally, when $\mu=0$, players are not influenced by their social environment, leading to the lowest average fitness across all game-based and strategy-based subpopulations. However, as $\mu$ increases, the average fitness of all players rises significantly. These findings demonstrate that the advantageous environment mechanism, enabled through the coupling parameter $\mu$, plays a crucial role in enhancing fitness across different social dilemmas and strategic interactions. Furthermore, increasing $\mu$ significantly improves the fitness of cooperators, thereby reinforcing cooperative behavior in the multi-game system.

\begin{table}[ht]
\scriptsize
\caption{Average fitness of cooperators and defectors in the entire system ($F_{C}$ and $F_{D}$), within $\Upsilon_{PDG}$ ($F_{PDG,C}$ and $F_{PDG,D}$), and within $\Upsilon_{SDG}$ ($F_{SDG,C}$ and $F_{SDG,D}$) for different values of {\it MCS}, considering $\sigma=0.05$, $\sigma=0.1$, $\sigma=0.5$, $\sigma=0.8$, and $\sigma=1.0$. The main control parameters are set to $b=1.10$ and $\mu=0.2$.}
\label{tablemcssigma}
\begin{tabularx}{\linewidth}{@{}LLLLLLLLLLLLL@{}}
\toprule
$\sigma$ & {\it MCS}    & 2      & 5      & 50     & 100    & 500    & 1000   & 5000   & 10000  & 50000  & 100000        \\ 
\midrule
$\sigma=0.05$   & {$F_{C}$}  &1.7838	&2.0410	&2.5489	&2.6269	&3.0644	&3.0758	&3.0437	&3.0297	&2.9955	&2.9822  \\
                & {$F_{SDG,C}$}      &1.8468	&2.1162	&2.6311	&2.7472	&3.1648	&3.1700	&3.1291	&3.1209	&3.0894	&3.0758  \\
                & {$F_{PDG,C}$}      &1.7176	&1.9488	&2.4309	&2.4676	&2.9466	&2.9638	&2.9401	&2.9188	&2.8807	&2.8675  \\
                & {$F_{D}$}  &1.4985	&0.8725	&0.3118	&0.3684	&0.9878	&0.9653	&0.9366	&0.9264	&0.9139	&0.9115  \\
                & {$F_{SDG,D}$}      &0.5129	&0.8577	&0.3022	&0.3556	&0.9367	&0.9090	&0.8841	&0.8771	&0.8663	&0.8659  \\
                & {$F_{PDG,D}$}      &1.4846	&0.8678	&0.3209	&0.3805	&1.0327	&1.0144	&0.9821	&0.9691	&0.9554	&0.9511 \\   
$\sigma=0.1$    & {$F_{C}$}  &1.8070	&2.1185	&2.5871	&2.7081	&3.1132	&3.1079	&3.0678	&3.0577	&3.0262	&3.0126  \\
                & {$F_{SDG,C}$}      &1.9364	&2.2638	&2.7651	&2.9020	&3.2691	&3.2491	&3.2058	&3.1926	&3.1611	&3.1490 \\
                & {$F_{PDG,C}$}      &1.6659	&1.9147	&2.3183	&2.4289	&2.8982	&2.9092	&2.8689	&2.8620	&2.8288	&2.8127  \\
                & {$F_{D}$}  &1.5057	&0.9421	&0.7054	&0.8461	&1.4809	&1.4130	&1.3559	&1.3368	&1.3100	&1.2945 \\
                & {$F_{SDG,D}$}      &1.5301	&0.9482	&0.6747	&0.7945	&1.3634	&1.3077	&1.2670	&1.2570	&1.2401	&1.2341  \\
                & {$F_{PDG,D}$}      &1.4827	&0.9331	&0.7317	&0.8890	&1.5581	&1.4820	&1.4146	&1.3898	&1.3571	&1.3355  \\
$\sigma=0.5$    & {$F_{C}$}  &2.0244	&2.6227	&3.2676	&3.4085	&3.4878	&3.4694	&3.4392	&3.4325	&3.4171	&3.4106  \\
                & {$F_{SDG,C}$}      &2.5804	&3.274	&3.6592	&3.7434	&3.7510	&3.7324	&3.7115	&3.7030	&3.6929	&3.6879  \\
                & {$F_{PDG,C}$}      &1.3241	&1.7503	&2.6978	&2.9453	&3.1035	&3.0745	&3.0191	&3.0114	&2.9846	&2.9744 \\
                & {$F_{D}$}  &1.5761	&1.5062	&2.2452	&2.4925	&2.3548	&2.2760	&2.1990	&2.1759	&2.1432	&2.1239  \\
                & {$F_{SDG,D}$}      &1.7043	&1.5870	&2.1576	&2.2756	&2.0333	&1.9921	&2.0171	&2.0068	&2.0169	&2.0040  \\
                & {$F_{PDG,D}$}      &1.4690	&1.4542	&2.2785	&2.5563	&2.4228	&2.3376	&2.2419	&2.2164	&2.1753	&2.1551  \\
$\sigma=0.8$    & {$F_{C}$}  &2.3135	&3.0782	&3.5270	&3.5927	&3.6332	&3.6257	&3.6092	&3.6063	&3.5953	&2.6906  \\
                & {$F_{SDG,C}$}      &3.1144	&3.6980	&3.9155	&3.9337	&3.9233	&3.9156	&3.9064	&3.9044	&3.8982	&3.8958  \\
                & {$F_{PDG,C}$}      &1.2211	&1.9976	&3.0124	&3.1550	&3.2321	&3.2131	&3.1726	&3.1658	&3.1408	&3.1300  \\
                & {$F_{D}$}  &1.5685	&1.7592	&2.6930	&2.8344	&2.5514	&2.4893	&2.4197	&2.3871	&2.3565	&2.3449  \\
                & {$F_{SDG,D}$}      &1.7341	&1.8509	&2.6592	&2.6202	&2.0755	&2.0215	&2.0647	&2.0939	&2.1352	&2.1515  \\
                & {$F_{PDG,D}$}      &1.4455	&1.7157	&2.7003	&2.8698	&2.6014	&2.5378	&2.4600	&2.4221	&2.3854	&2.3711  \\ 
$\sigma=1.0$    & {$F_{C}$}  &2.4321	&3.1863	&3.5599	&3.6625	&3.7512	&3.7508	&3.7515	&3.7458	&3.7455	&3.7462  \\
                & {$F_{SDG,C}$}      &3.4139	&3.9390	&3.9973	&4.0000	&4.0000	&4.0000	&4.0000	&4.0000	&4.0000	&4.0000  \\
                & {$F_{PDG,C}$}      &1.0599	&1.8619	&2.9882	&3.2499	&3.4616	&3.4586	&3.4582	&3.4440	&3.4420	&3.4422  \\
                & {$F_{D}$}  &1.5533	&1.8405	&2.6609	&2.6988	&2.4816	&2.4591	&2.4473	&2.4342	&2.4497	&2.4202  \\
                & {$F_{SDG,D}$}      &1.7457	&1.9788	&2.6206	&2.4860	&2.2135	&2.2425	&2.1423	&2.0818	&2.1011	&2.0056  \\
                & {$F_{PDG,D}$}      &1.1455	&1.7804	&2.6691	&2.3442	&2.5232	&2.4913	&2.4916	&2.4839	&2.4986	&2.4774  \\ 
\bottomrule
\end{tabularx}
\end{table}

Now, we investigate the underlying mechanisms by which the sucker’s payoff $\sigma$ influences different strategies and dilemmas, Table~\ref{tablemcssigma} presents the average fitness of cooperators and defectors across the entire system, as well as within $\Upsilon_{PDG}$ and $\Upsilon_{SDG}$, as a function of the evolutionary time steps for $\sigma=0.05$, $\sigma=0.1$, $\sigma=0.5$, $\sigma=0.8$, and $\sigma=1.0$. On the one hand, throughout the entire evolutionary process, regardless of the number of time steps, an increase in $\sigma$ leads to a significant rise in the average fitness of both cooperators and defectors across the entire system, $\Upsilon_{PDG}$, and $\Upsilon_{SDG}$. Moreover, in all cases, the average fitness of cooperators remains consistently higher than that of defectors. On the other hand, for any given value of $\sigma$, in the early time steps, the average fitness of both defectors and cooperators is higher within $\Upsilon_{SDG}$ compared to the overall system. However, the average fitness of defectors in $\Upsilon_{PDG}$ exceeds that of defectors in $\Upsilon_{SDG}$. This phenomenon can be understood in the context of social dilemmas. The dilemma within $\Upsilon_{PDG}$ is more severe than in $\Upsilon_{SDG}$, meaning that cooperation is more challenging to sustain in the PDG subpopulation. Consequently, defectors in $\Upsilon_{PDG}$ and cooperators in $\Upsilon_{SDG}$ obtain higher payoffs compared to their counterparts in the opposite dilemma setting.

\begin{figure}
    \centering
    \includegraphics[width=1.0\textwidth]{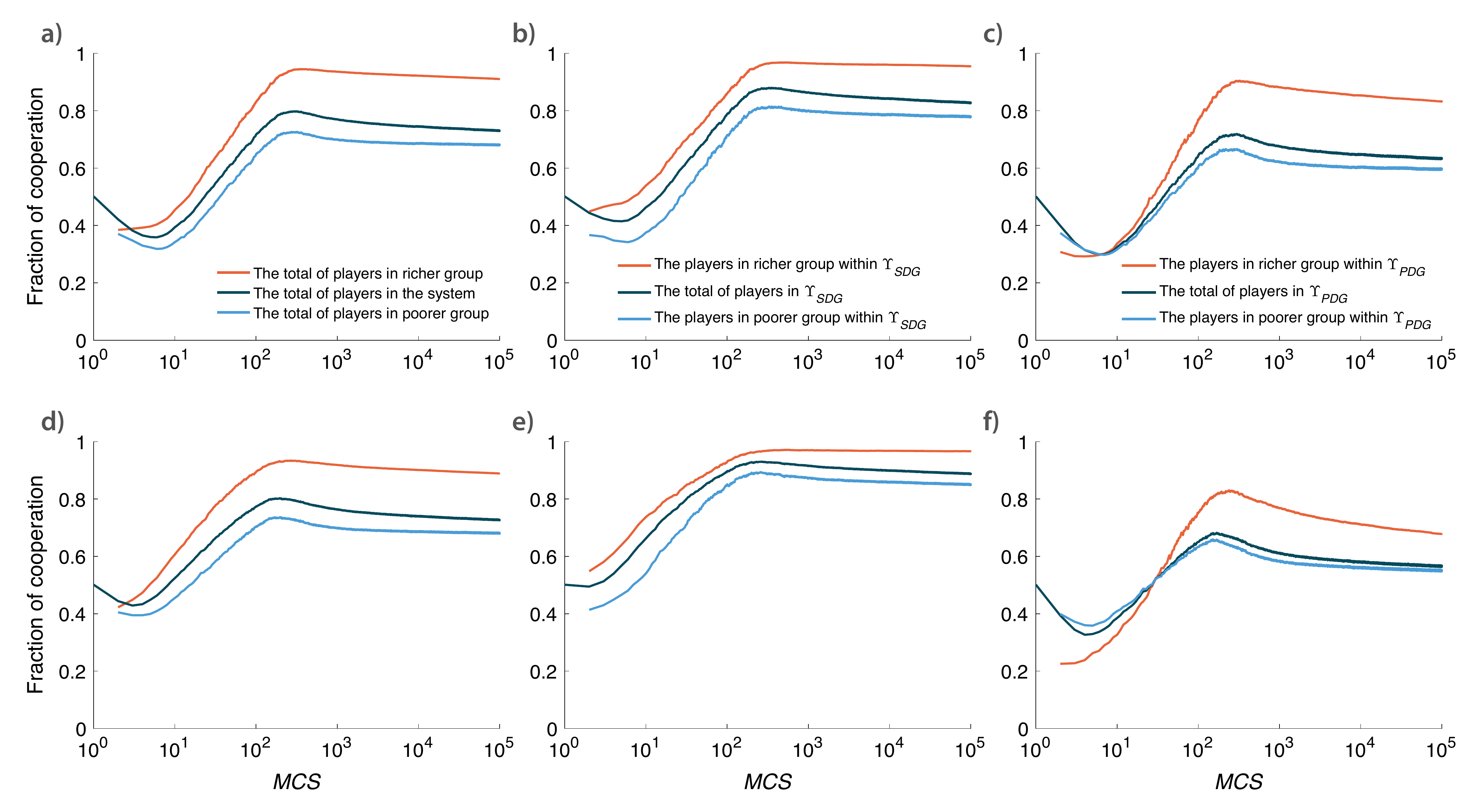}
    \caption{Time evolution in Monte Carlo Steps (MCS) of the fraction of cooperation, distinguishing individuals based on resource quantity and game type. Two main groups are considered: the richer group (individuals with resources greater than 1 unit) and the poorer group (individuals with resources less than 1 unit). Left column (panels (a) and (d)): Evolution of the fraction of cooperation for the entire system, as well as for the richer and poorer groups. Central column (panels (b) and (e)) and right column (panels (c) and (f)) presents the corresponding evolution within $\Upsilon_{SDG}$ and $\Upsilon_{PDG}$, again distinguishing between the richer and poorer groups, respectively. The top row (panels (a)--(c)) corresponds to the parameter set $\sigma=0.2$, $\mu=0.5$, while the bottom row (panels (d)--(f)) corresponds to $\sigma=0.5$, $\mu=0.2$. The remaining control parameter is set to $b=1.10$.}
    \label{figwtheat9}
\end{figure}

To further analyze the influence of limited resources on the evolution of cooperation in this multi-game model, we examine the relationship between players’ resource quantities, game-based subpopulations, and strategic choices as a function of $\mu$ and $\sigma$. Based on the quantity of resources owned, players are divided into two groups: the richer group (individuals with more than one unit of resources) and the poorer group (individuals with less than one unit of resources).  

The first row in Fig.~\ref{figwtheat9} presents the time evolution, in {\it MCS}, of the fraction of cooperation in the entire system, as well as in the richer and poorer groups, when $\mu=0.5$ and $\sigma=0.2$. The results for additional values of $\mu=0$, $0.2$, $0.8$, $1.0$ are shown in the Appendix, Fig.~\ref{figa1}. In Fig.~\ref{figwtheat9}(a) and the first column of Fig.~\ref{figa1}, the red line represents the cooperation levels in the richer group, while the blue line represents the cooperation levels in the poorer group. It can be observed that when $\mu=0$, $\mu=0.2$, and $\mu=0.5$, the cooperation fraction initially decreases, reaching its minimum after approximately $5$ {\it MCS}. Subsequently, cooperation increases until reaching a peak at around $100$ {\it MCS}, after which it stabilizes. The overall trend for the poorer group follows a similar pattern but remains consistently lower in magnitude. When the system reaches a steady state, the cooperation levels in the richer group are about $1.2$ times those of the poorer group.  In Fig.~\ref{figwtheat9}(b) and Fig.~\ref{figwtheat9}(c), as well as the second and third columns in Fig.~\ref{figa1}, show the evolution of the fraction of cooperation in the richer and poorer groups over time within $\Upsilon_{SDG}$ and $\Upsilon_{PDG}$, respectively. The overall trends are consistent with those observed for the entire system. However, in the early stages of evolution, both richer and poorer groups in $\Upsilon_{PDG}$ exhibit lower cooperation levels than in $\Upsilon_{SDG}$ and the overall system. Notably, in $\Upsilon_{PDG}$, the fraction of cooperation in the poorer group initially surpasses that of the richer group by a significant margin until it reaches its minimum value. This phenomenon can be attributed to the nature of the social dilemmas and the random endorsement of strategies. At the beginning of the simulation, strategies are randomly distributed within $\Upsilon_{SDG}$ and $\Upsilon_{PDG}$, which initially benefits defectors. Furthermore, since $\Upsilon_{PDG}$ is characterized by a stronger dilemma compared to $\Upsilon_{SDG}$, the minimum cooperation level in $\Upsilon_{PDG}$ is lower than in $\Upsilon_{SDG}$. Additionally, cooperators in $\Upsilon_{PDG}$ face a significant disadvantage, making them more susceptible to strategy replacement and resource loss. As a result, cooperative behavior in $\Upsilon_{PDG}$ remains at a lower level. On the other hand, as $\mu$ increases, this disadvantage is mitigated, improving cooperative behavior throughout the evolutionary process in the entire system, $\Upsilon_{SDG}$, and $\Upsilon_{PDG}$. Through this analysis, it can be concluded that increasing $\mu$ significantly reduces the initial disadvantage of cooperators, shortens the time required for the system to reach a stable state, and enhances cooperation across the system, including within $\Upsilon_{SDG}$ and $\Upsilon_{PDG}$. 

Now, the bottom row of Fig.~\ref{figwtheat9} presents the cooperation fraction in the richer and poorer groups over time for the entire system, $\Upsilon_{SDG}$, and $\Upsilon_{PDG}$ when $\sigma=0.5$ and $\mu=0.2$. The corresponding results for additional values of $\sigma=0.05$, $0.1$, $0.8$, $1.0$ are shown in Fig.~\ref{figa2} in the Appendix. It can be observed that for small values of $\sigma$, throughout the entire evolutionary process, the cooperation ratio in the richer and poorer groups follows the same pattern as in the case of small $\mu$: it initially decreases, reaches a minimum, and then continuously increases until the system reaches a stable state. As $\sigma$ increases, cooperation is significantly enhanced across the system and within the groups in $\Upsilon_{SDG}$. However, in $\Upsilon_{PDG}$, when $\sigma$ is high, the cooperation fraction in the poorer group exceeds that in the richer group throughout the entire evolution process. This phenomenon can be explained in terms of social dilemmas. When $\sigma$ is large, cooperators in $\Upsilon_{SDG}$ and defectors in $\Upsilon_{PDG}$ benefit from more favorable conditions. In contrast, cooperators in $\Upsilon_{PDG}$ face greater difficulty in spreading their strategy to other individuals to acquire resources. Consequently, they are more likely to be replaced by other strategies and lose resources. As a result, in $\Upsilon_{PDG}$, the cooperation fraction remains lower in the richer group than in the poorer group.  
These findings confirm that $\sigma$ plays a crucial role in shaping cooperative behavior within the richer and poorer groups of both $\Upsilon_{SDG}$ and $\Upsilon_{PDG}$, as well as influencing the overall resource distribution across individuals in different game settings.

\begin{figure}
    \centering
    \includegraphics[width=1.0\textwidth]{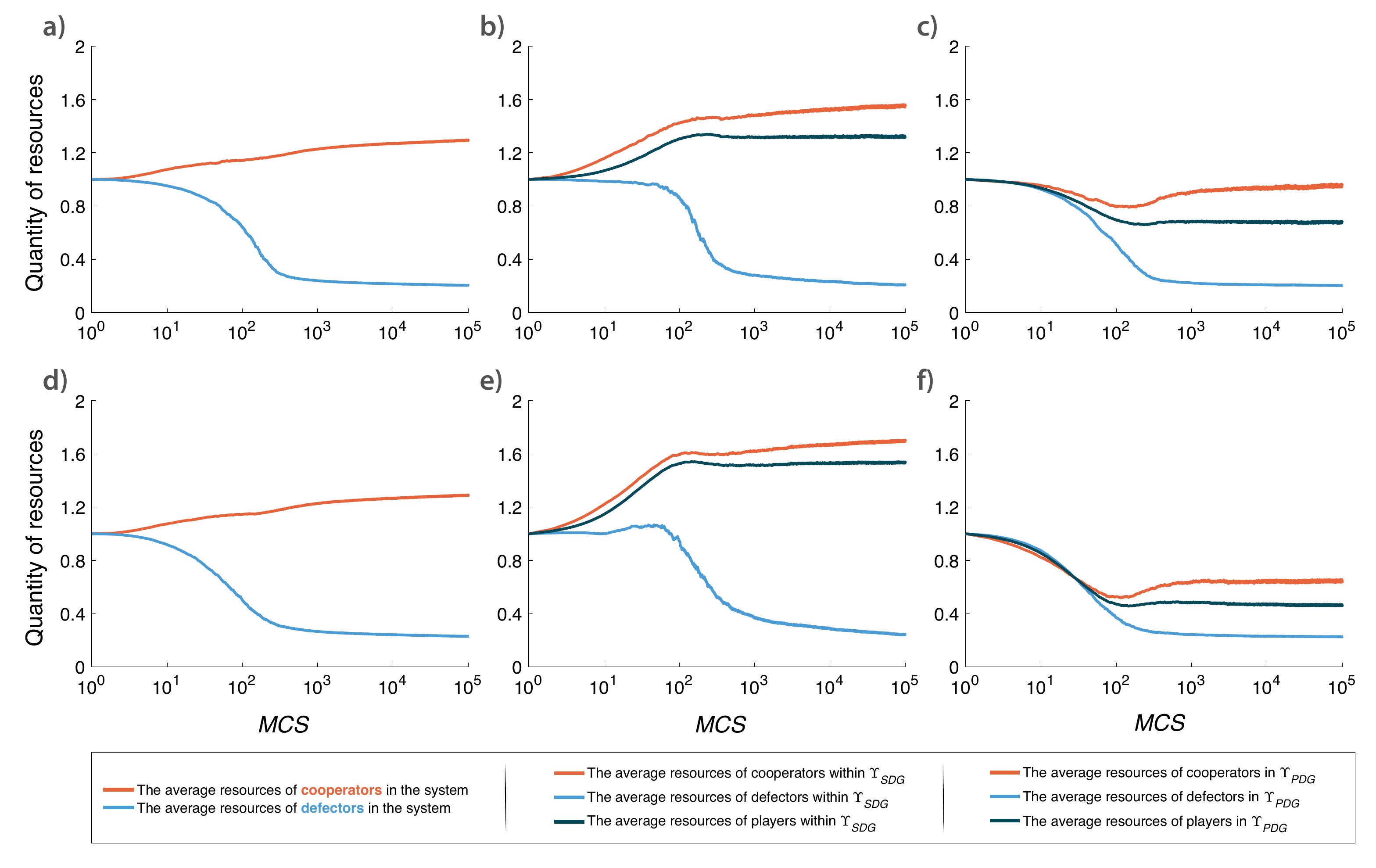}
    \caption{Time evolution in Monte Carlo Steps (MCS) of the average resources of cooperators and defectors across the entire system and within $\Upsilon_{SDG}$ and $\Upsilon_{PDG}$. The left column (panels (a) and (d)) shows the system-wide evolution of average resources for both cooperators and defectors. The central column (panels (b) and (e)) focuses on the same observable within $\Upsilon_{SDG}$, while the right column (panels (c) and (f)) presents the corresponding evolution within $\Upsilon_{PDG}$. The top row (panels (a)--(c)) corresponds to the parameter set $\sigma=0.2$, $\mu=0.5$, whereas the bottom row (panels (d)--(f)) corresponds to $\sigma=0.5$, $\mu=0.2$. The remaining control parameter is set to $b=1.10$.}
    \label{figwtheat10}
\end{figure}

We continue by analyzing how resources are distributed across strategy-based and game-based subpopulations and this is impacted by
the coupling parameter $\mu$ and the sucker’s payoff $\sigma$. The first row in Fig.~\ref{figwtheat10} shows the evolution of the average resources for cooperators and defectors in the entire system, as well as within $\Upsilon_{SDG}$ and $\Upsilon_{PDG}$, as a function of time, for $\sigma=0.2$ and $\mu=0.5$. The corresponding results for other values of $\mu$ ($\mu=0$, $0.2$, $0.8$, $1.0$) are shown in Fig.~\ref{figa3} (Appendix).  
In Fig.~\ref{figwtheat10}(a) and the first column of Fig.~\ref{figa3}, the evolution of the mean resources of cooperators and defectors is illustrated. It can be observed that when $\mu=0$, $0.2$, $0.5$, the mean quantity of resources of cooperators initially increase before stabilizing, while the mean resources of defectors initially decrease before reaching a steady state. At all stages of the evolution, cooperators consistently maintain higher resource levels than defectors. As $\mu$ increases, the magnitude of change in cooperators’ average resources decreases, leading to a reduced gap between cooperators and defectors. The second and third columns of Fig.~\ref{figwtheat10} and Fig.~\ref{figa3} show the evolution of the average resources of cooperators and defectors in $\Upsilon_{SDG}$ and $\Upsilon_{PDG}$, respectively. The trends observed in $\Upsilon_{SDG}$ closely follow those seen in the entire system. Additionally, for $\mu=0$, $0.2$, $0.5$, cooperators and defectors in $\Upsilon_{SDG}$ maintain average resource levels above 1, whereas in $\Upsilon_{PDG}$, the average resource levels remain below $1$. At any fixed value of $\mu$, the average resources in $\Upsilon_{SDG}$ are consistently higher than those in $\Upsilon_{PDG}$. However, as $\mu$ increases, the average resources in $\Upsilon_{PDG}$ increase, while those in $\Upsilon_{SDG}$ slightly decrease, narrowing the resource gap between the two subpopulations. These findings suggest that the resource availability within different game types significantly influences cooperation rates, with wealthier individuals exhibiting a stronger tendency toward cooperation. Furthermore, $\mu$ not only fosters cooperation across the system but also mitigates resource inequality between different social dilemmas and strategies.

The second row in Fig.~\ref{figwtheat10} shows the evolution of the mean resources of cooperators and defectors in the entire system, $\Upsilon_{SDG}$, and $\Upsilon_{PDG}$ for $\mu=0.2$ and $\sigma=0.5$. The corresponding results for other values of $\sigma$ ($\sigma=0.05$, $0.1$, $0.8$, $1.0$) are presented in Fig.~\ref{figa4} (Appendix). It can be observed that when $\sigma$ is small ($\sigma=0.05$, $0.1$, $0.5$), cooperators maintain significantly higher resource levels than defectors across the entire system, $\Upsilon_{SDG}$, and $\Upsilon_{PDG}$, which is consistent with earlier findings. As $\sigma$ increases, the average resources of cooperators in $\Upsilon_{SDG}$ decline, while defectors’ resources increase. However, in $\Upsilon_{PDG}$, the average resources of both cooperators and defectors decline continuously. This indicates that as $\sigma$ increases, individuals in $\Upsilon_{PDG}$ experience stronger dilemmas, making them more vulnerable to being replaced by strategies from individuals in $\Upsilon_{SDG}$. These results demonstrate that $\sigma$ plays a critical role in determining the resource distribution among individuals adopting different strategies. While a higher $\sigma$ enhances resource acquisition for individuals in $\Upsilon_{SDG}$, it also exacerbates the disparity in resource distribution between $\Upsilon_{SDG}$ and $\Upsilon_{PDG}$.

\begin{table}[ht]
\scriptsize
\caption{The variables $N$ ($N_{SDG}$, $N_{PDG}$), $r$ ($r_{SDG}$, $r_{PDG}$), and $f_C$ ($f_{SDG,C}$, $f_{PDG,C}$) represent the number of players, total resource quantity, and cooperation fraction, respectively, for the entire system, $\Upsilon_{SDG}$, and $\Upsilon_{PDG}$. Results for $b=1.05$, $b=1.10$, $b=1.30$, and $b=1.50$ and disaggregated by resource intervals. Other parameters are $\mu=0.5$, $\sigma=0.5$.}
\label{tablemressource}
\begin{tabularx}{\linewidth}{@{}LLLLLLLLLLLLL@{}}
\toprule
$b$ & {\it Resources}    & (0,0.1)    & (0.1,0.2)      & (0.2,0.5)     & (0.5,0.8)    & (0.8,1.1)     & (1.1,1.4)   & (1.4,1.7)   & (1.7,2.0)  & (2.0,5.0)  &(5.0,10.0) &$>10.0$       \\ 
\midrule
$b=1.10$        & {$N$}   &29051	&603	&445	&562	&461	&408	&374	&351	&4623	    &3039	    &83 \\
                & {$N_{SDG}$}       & 13325	&82	    &86	    &112  	&126	&160	&157	&187	&3436	    &2287	    &65 \\
                & {$N_{PDG}$}       &15626	&521	&359	&450	&335	&248	&217	&164	&1187	    &752	    &18   \\
                & {$r$}   &30.43	&116.05	&176.77	&357.50	&437.83	&511.93	&580.56	&647.10	&16568.83	&19644.78	&928.19   \\
                & {$r_{SDG}$}       &5.08	&15.86	&34.28	&73.55	&119.48	&204.05	&245.06 &344.51 &12420.37	&14759.85	&726.08  \\
                & {$r_{PDG}$}       &25.35	&100.20	&142.49	&283.95	&318.35	&307.88	&335.50	&302.60	&4148.46	&4884.92	&202.11  \\   
                & {$f_C$}   &0.7607	&0.2769	&0.2674	&0.3256	&0.3926	&0.4828	&0.5989	&0.7009	&0.9701  	&0.9993	    &1.0000  \\
                & {$f_{SDG,C}$}       &0.8878	&0.5244	&0.6047	&0.6875	&0.6746	&0.7937	&0.7962	&0.8770	&0.9860	    &0.9996	    &1.0000 \\
                & {$f_{PDG,C}$}       &0.6530	&0.2380	&0.1866	&0.2356	&0.2866	&0.2823	&0.4562	&0.5000	&0.9242	    &0.9987	    &1.0000  \\                
$b=1.30$        & {$N$}   &30575	&867	&625	&857	&647	&547	&475	&337	&1770	    &2270	    &1030  \\
                & {$N_{SDG}$}       &14221	&248	&170	&260	&231	&207	&223	&176	&1331	    &2009	    &947  \\
                & {$N_{PDG}$}       &16354	&619	&455	&597	&416	&340	&252	&1661	&439	    &261	    &83  \\
                & {$r$}   &45079	&167.42	&249.27	&548.58	&614.67	&681.15	&739.09	&621.84	&6028.94	&16419.38	&13883.83   \\
                & {$r_{SDG}$}       &12.93	&46.93	&67.94	&167.20	&221.47	&259.41	&349.26	&324.45	&4680.05	&14587.63	&12784.32\\
                & {$r_{PDG}$}       &32.86	&120.49	&181.33	&381.38	&393.20	&421.74	&389.83	&297.39	&1348.89	&1831.74	&1099.51  \\   
                & {$f_C$}   &0.4835	&0.1522	&0.1408	&0.1190	&0.1762	&0.2176	&0.2779	&0.3591	&0.7215	    &0.9740	    &0.9854  \\
                & {$f_{SDG,C}$}       &0.6689	&0.3589	&0.3471	&0.2885	&0.4069	&0.4493	&0.4978	&0.5852	&0.8475	    &0.9816	    &0.9884 \\
                & {$f_{PDG,C}$}       &0.3222	&0.0695	&0.0637 &0.0452	&0.0481	&0.0765	&0.0833	&0.1118	&0.3394  	&0.9157	    &0.9518  \\  
$b=1.50$        & {$N$}   &30708	&972	&705	&901	&744	&645	&509	&441	&1661	    &1376	    &1338   \\
                & {$N_{SDG}$}       &14545	&353	&265	&340	&285	&299	&260	&214	&1061	    &1177	    &1224 \\
                & {$N_{PDG}$}       &16163	&619	&440	&561	&459	&346	&249	&227	&600	    &199	    &114  \\
                & {$r$}   &47.67	&187.60	&278.77	&580.20	&701.89	&802.22	&784.93	&812.43	&5214.79	&10092.84	&20496.61   \\
                & {$r_{SDG}$}       &17027	&67.96	&104.65	&219.69	&269.67	&373.00	&400.33	&393.59	&3473.92	&8711.68	&18851.91    \\
                & {$r_{PDG}$}       &30.40	&119.64	&174.12	&360.51	&432.21	&429.22	&384.60	&418.84	&1740.87	&1381.16	&1644.70    \\   
                & {$f_C$}   &0.3278	&0.1163	&0.0993	&0.1021	&0.1156	&0.1473	&0.1749	&0.2063	&0.3715	    &0.8365  	&0.9387    \\
                & {$f_{SDG,C}$}       &0.4704	&0.2266	&0.1962	&0.2294	&0.2561	&0.2742	&0.3154	&0.3832	&0.5316     &0.8921 	&0.9567    \\
                & {$f_{PDG,C}$}       &0.1995	&0.0533	&0.0409	&0.0250	&0.0283	&0.0376	&0.0281	&0.0396	&0.0883	    &0.5075	    &0.7456    \\  
$b=1.70$        & {$N$}   &29372	&831	&705	&801	&639	&576	&505	&467	&3522	    &1877	    &705    \\
                & {$N_{SDG}$}      &13713	&368	&310	&341	&299	&279	&250	&240	&2181	    &1390	    &652      \\
                & {$N_{PDG}$}      &15659	&463	&395	&460	&340	&297	&255	&227	&1341	    &487	    &53      \\
                & {$r$}  &35.38	&159.50	&279.69	&510.79	&605.09	&718.76	&782.46	&836.60	&11751.09	&12703.83	&11589.79     \\
                & {$r_{SDG}$}      &17.51	&71.40	&122.64	&215.72	&284.16	&347.37	&388.19	&443.06	&7416.60	&9555.63	&10760.04  \\
                & {$r_{PDG}$}      &17.88	&88.10	&157.05	&295.07	&320.92	&371.34	&394.27	&420.53	&4334.49	&3148.20	&829.75  \\   
                & {$f_C$}  &0.1201	&0.0385	&0.0298	&0.0312	&0.0360	&0.0278	&0.0356	&0.0404	&0.0542	    &0.2200   	&0.7603    \\
                & {$f_{SDG,C}$}      &0.1792	&0.0652	&0.0548	&0.0587	&0.0702	&0.0538	&0.0720	&0.0792	&0.0830  	&0.2878 	&0.8052   \\
                & {$f_{PDG,C}$}      &0.0683	&0.0173	&0.0101	&0.0109	&0.0059	&0.0034	&0.0000	&0.0000	&0.0075 	&0.0267	    &0.2075   \\  
\bottomrule
\end{tabularx}
\end{table}

To further examine how individuals' games and strategy choices interplay with resource availability, we classify individuals into $11$ groups based on their resource levels once the system stabilizes. The total number of individuals, resource levels, and cooperation fractions for the entire system, $\Upsilon_{SDG}$, and $\Upsilon_{PDG}$ are presented in Table~\ref{tablemressource}. Regardless of the value of $b$, we observe that individuals in $\Upsilon_{PDG}$ are predominantly concentrated in lower resource intervals (below $1.7$), whereas individuals in $\Upsilon_{SDG}$ are primarily found in higher resource intervals. Furthermore, the individuals' resource distribution is highly imbalanced. In the lowest resource interval, approximately $60\%$ of individuals are concentrated, yet they collectively hold only about $1\%$ of the total system resources. As resource levels increase, the cooperation fraction rises across the entire system, as well as within $\Upsilon_{PDG}$ and $\Upsilon_{SDG}$. At any given resource level, the cooperation fraction is highest in $\Upsilon_{SDG}$ and lowest in $\Upsilon_{PDG}$. Additionally, $b$ has a pronounced effect on the cooperation fraction across different resource intervals. As $b$ increases, cooperation levels decline within the same resource interval. This analysis further confirms that cooperative behaviors in different dilemmas are closely linked to resource availability, with resources playing a crucial role in shaping the behavior of different types of cooperators. Moreover, $\Upsilon_{SDG}$ exhibits a stronger capacity for resource acquisition, reinforcing previous findings.

\begin{figure}
    \centering
    \includegraphics[width=1.0\textwidth]{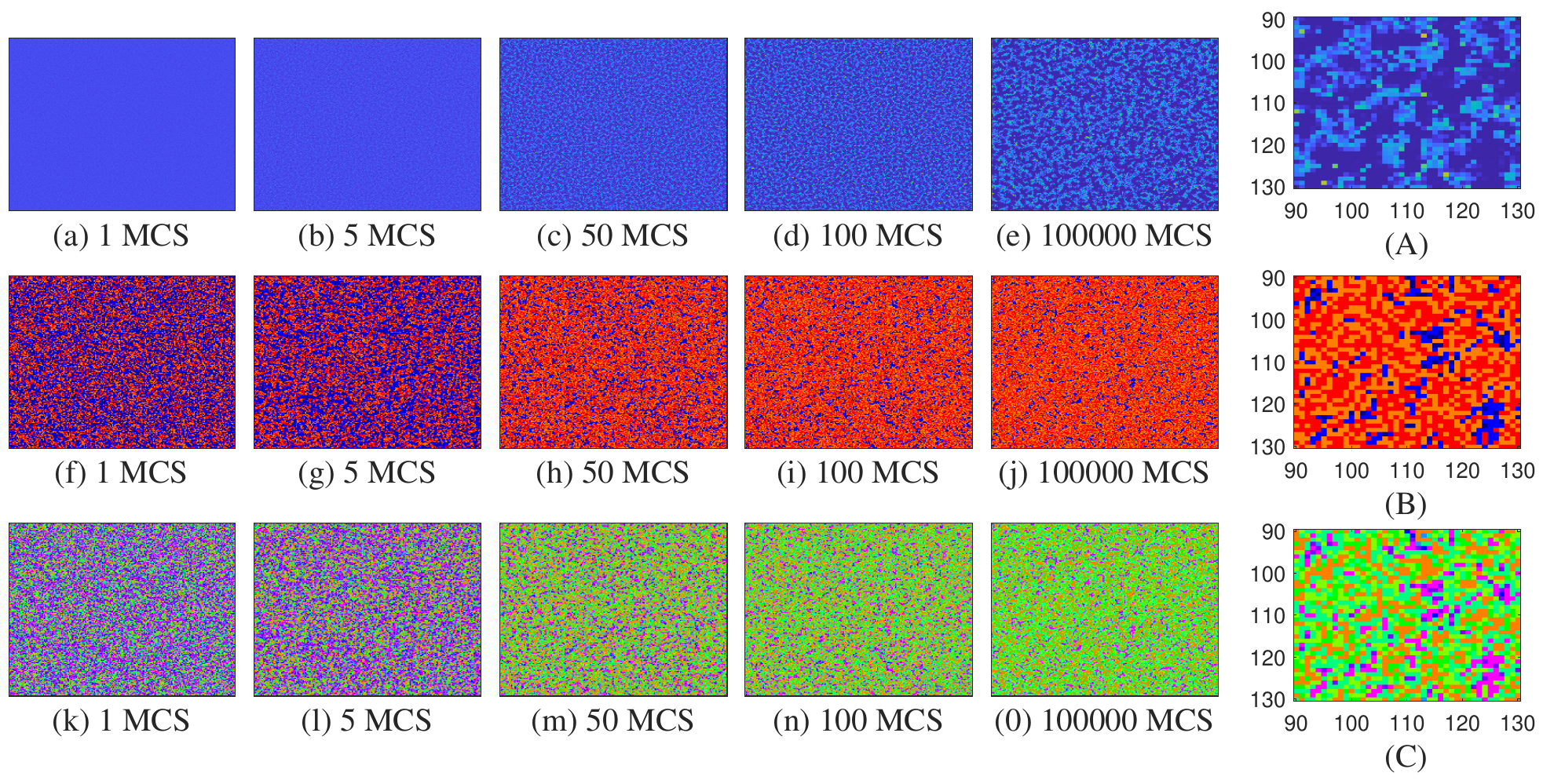}
    \caption{Characteristic configuration snapshots of the spatial multi-game framework at different {\it MCS} steps. The first row (panels (a)--(e)) presents the evolution of resource distribution, where darker colors indicate lower resource levels. Panel (A) is a magnified section of panel (e). The second row (panels (f)--(j)) shows the evolution of strategy distribution within $\Upsilon_{SDG}$ and $\Upsilon_{PDG}$. Panel (B) is a magnified section of panel (j). Defectors and cooperators are represented in blue and red for $\Upsilon_{SDG}$, and in dark blue and orange for $\Upsilon_{PDG}$. The third row (panels (k)--(o)) illustrates the co-evolution of strategies and resources in $\Upsilon_{SDG}$ and $\Upsilon_{PDG}$. Panel (C) is a magnified section of panel (o). Within $\Upsilon_{SDG}$, cooperators and defectors are represented in orange and dark blue, respectively. Within $\Upsilon_{PDG}$, cooperators and defectors are represented in pure green and dodger blue, yellow-green and cyan, and purple and magenta. Columns from left to right correspond to {\it MCS} steps $1$, $5$, $50$, $100$, and $100000$. The remaining control parameters are set to $\sigma=0.5$, $\mu=0.5$, and $b=1.10$.}
    \label{figwtheat11}
\end{figure}

To gain deeper insights into the evolution of resource distribution and its interaction with the different games and strategies, we analyze characteristic snapshots of individuals' resources, strategy dynamics, and their coevolution over time, as shown in Fig.~\ref{figwtheat11}.  

The first row (panels (a)--(e)) illustrates the evolution of resource distribution. Initially, each player is assigned one unit of resources, and resource dynamics unfold rapidly. By approximately $100$ {\it MCS} (panels (b)--(d)), it becomes evident that resource distribution is highly uneven. Once the system stabilizes, as shown in Fig.~\ref{figwtheat11}(e), the disparity is stark. Fig.~\ref{figwtheat11}(A) provides a magnified view of a localized region in Fig.~\ref{figwtheat11}(e), where dark regions (indicating individuals with fewer resources) dominate most of the system, while bright spots (individuals with higher resources) are sparsely scattered, confirming the strong heterogeneity in resource redistribution. The second row (panels (f)--(j)) presents the evolution of strategy configuration within $\Upsilon_{SDG}$ and $\Upsilon_{PDG}$. Here, defectors are represented in dark blue and blue for $\Upsilon_{SDG}$ and $\Upsilon_{PDG}$, respectively, while cooperators are marked in red and orange for $\Upsilon_{SDG}$ and $\Upsilon_{PDG}$. At the beginning of the evolution, cooperators and defectors are randomly distributed across the system. As the system evolves, the number of blue and dark blue nodes first increases (panels (f) and (g)) before gradually declining, while red and orange nodes proliferate, forming cooperative clusters. Upon stabilization, as seen in Fig.~\ref{figwtheat11}(j), cooperators in $\Upsilon_{SDG}$ (red) and $\Upsilon_{PDG}$ (orange) form large cooperative clusters, occupying most of the system, while defectors (blue and dark blue) become sparse and scattered. Fig.~\ref{figwtheat11}(B) provides a magnified view of Fig.~\ref{figwtheat11}(j), highlighting how cooperators in $\Upsilon_{PDG}$ (orange) are predominantly positioned at the periphery of the cooperative clusters. This spatial arrangement facilitates resource acquisition by $\Upsilon_{PDG}$ cooperators from defectors, while also benefiting $\Upsilon_{SDG}$ cooperators by allowing them to access resources from those in $\Upsilon_{PDG}$. The third row (panels (k)--(o)) shows the coevolution of strategies and resources, distinguishing between richer and poorer groups within $\Upsilon_{SDG}$ and $\Upsilon_{PDG}$. In the early stages of evolution, the number of defectors (represented in purple, dark blue, magenta, and dodger blue) initially increases before gradually declining, while cooperators (yellow-green, orange, blue-green, and orange) follow the opposite trend, forming increasingly stable cooperative clusters. Once the system reaches equilibrium, as shown in Fig.~\ref{figwtheat11}(o), cooperators in richer groups within $\Upsilon_{SDG}$ (orange) are primarily located inside cooperative clusters, while cooperators in poorer groups within $\Upsilon_{PDG}$ (yellow-green) are mostly positioned along cluster borders. Fig.~\ref{figwtheat11}(C) provides a magnified view of Fig.~\ref{figwtheat11}(o), illustrating how this spatial distribution benefits cooperators: those at the cluster borders can extract resources from nearby defectors, while those within richer groups in $\Upsilon_{SDG}$ can reduce their own resource losses and gain additional resources from other individuals.

\begin{figure}[h]
    \centering
    \makebox[\textwidth][l]{\hspace{-0.12\textwidth}
        \begin{adjustbox}{scale=1.2} % Increase this to scale everything uniformly
            \begin{minipage}{\textwidth}
                \begin{subfigure}{0.55\textwidth}
                    \centering
                    \includegraphics[width=\linewidth, height=4cm]{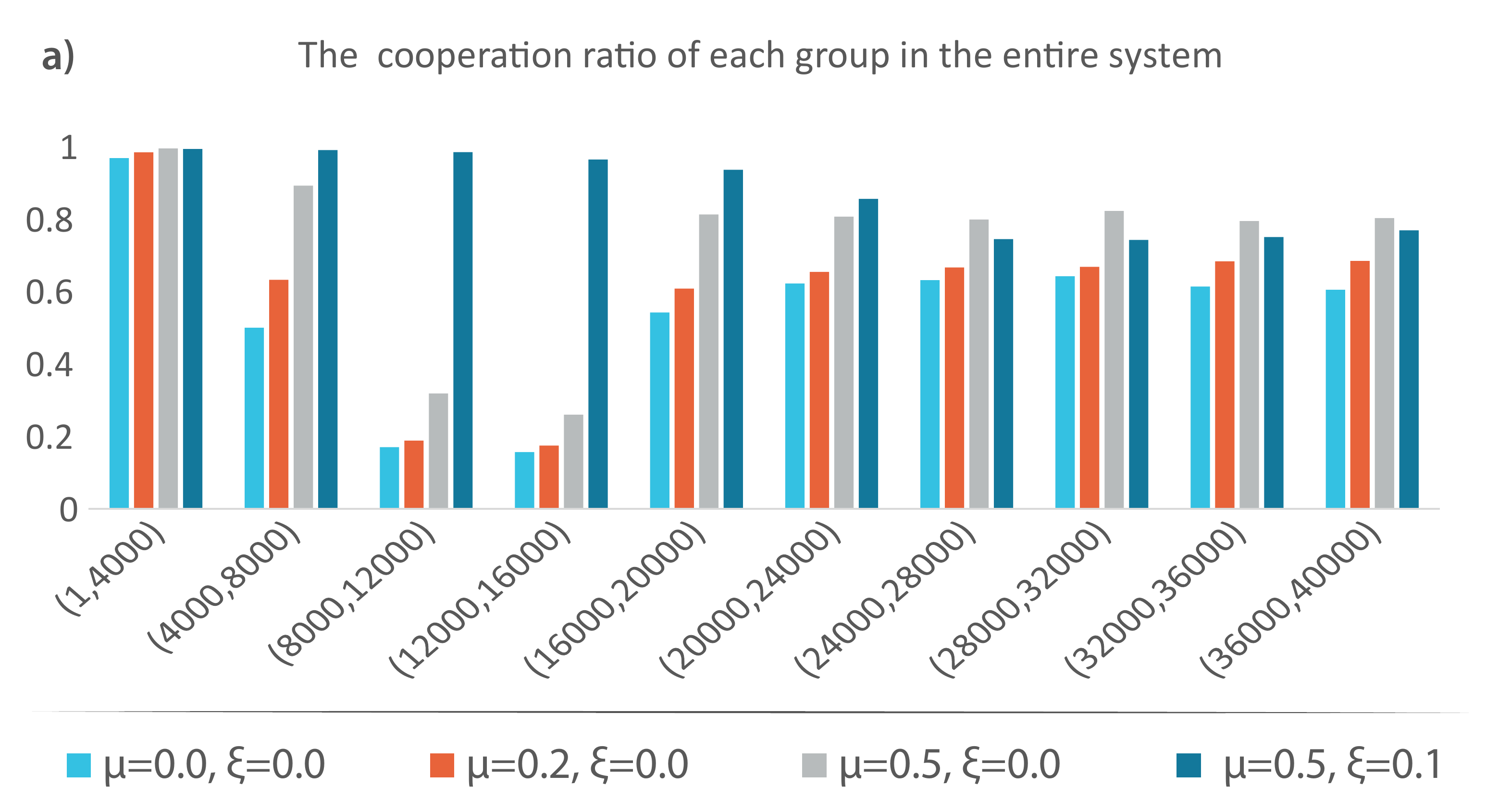}
                \end{subfigure}
                \hspace{0.02\textwidth}
                \begin{subfigure}{0.30\textwidth}
                    \centering
                    \includegraphics[width=\linewidth]{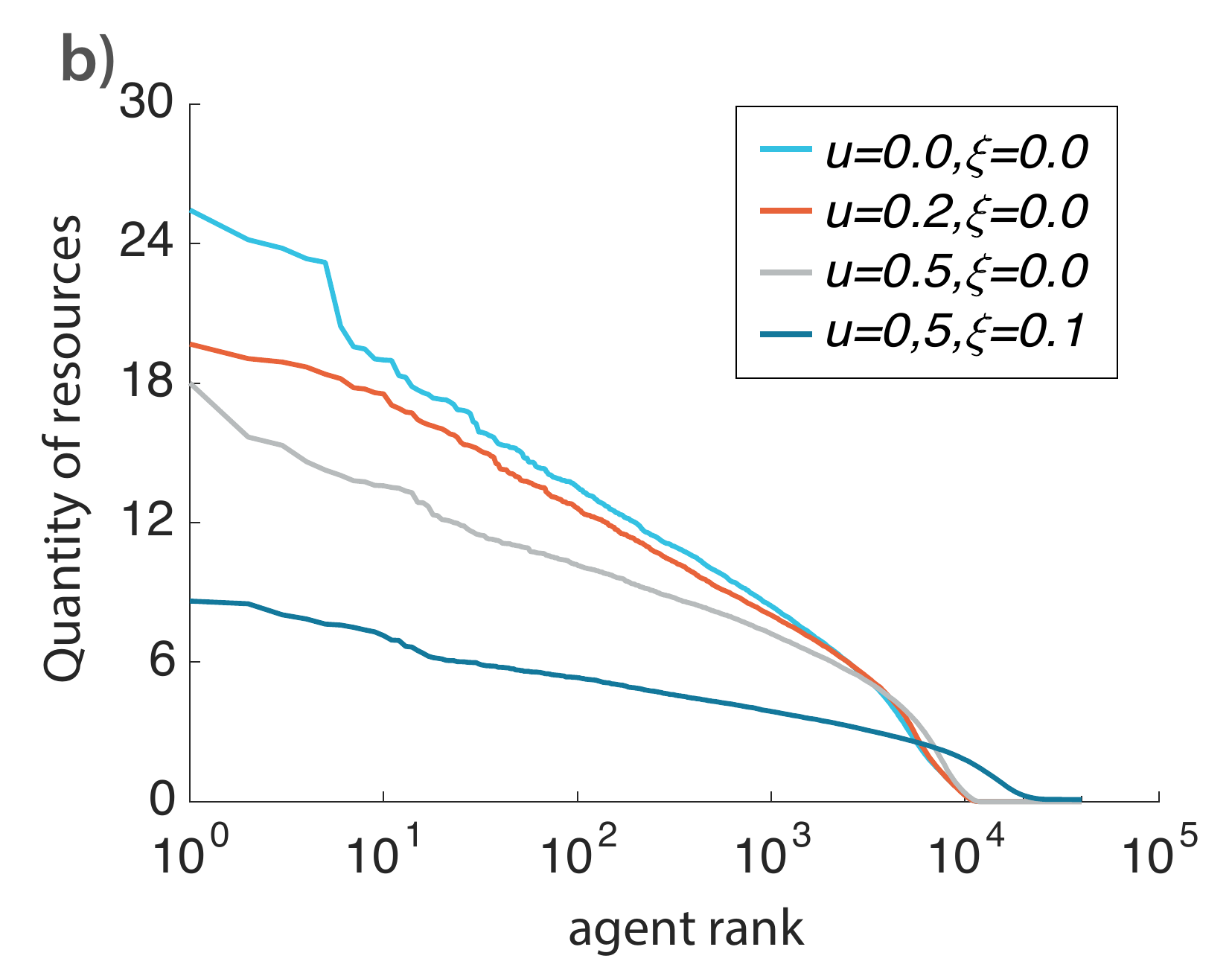}
                \end{subfigure}
            \end{minipage}
        \end{adjustbox}
    }
    \caption{Analysis of the relationship between $\mu$, resource quantity, and cooperation fraction when the system reaches equilibrium. Players in the entire system, $\Upsilon_{SDG}$, and $\Upsilon_{PDG}$ are evenly divided into ten groups based on their resource levels. Left column (panels (a) and (b)): Cooperation fraction and corresponding resource distribution across different resource-based groups for the entire system. Right column (panels (c) and (d)): Cooperation fraction and resource rank-plot for $\Upsilon_{SDG}$ and $\Upsilon_{PDG}$, respectively. Panel (a) presents the cooperation fraction for different values of $\mu$, specifically $\mu=0$, $\mu=0.2$, $\mu=0.5$, and $\mu=0.5$ with $\xi=0.1$, where $\xi$ represents the resource guarantee parameter. The corresponding resource rank-plot for the system is shown in panel (b). The same analysis is extended to $\Upsilon_{SDG}$ and $\Upsilon_{PDG}$ in Fig.~\ref{figa5}(a)--(b) and Fig.~\ref{figa5}(c)--(d), respectively. The remaining control parameters are set to $\sigma=0.2$ and $b=1.10$.}
    \label{figwtheat12}
\end{figure}

Based on the earlier analysis, it is evident that the system's resource distribution is highly uneven. To further investigate the relationship between resource availability, cooperation fraction, and the advantageous environment, as enabled by $\mu$, across the entire system and within both games, as well as to mitigate resource inequality among individuals, we divide the population into ten groups, ranked from highest to lowest, based on individual resource levels.

Fig.~\ref{figwtheat12}(a) and Fig.~\ref{figwtheat12}(b) illustrate the relationship between cooperation fraction and resource levels across different resource intervals, as well as the individual resource rank-plot for different values of $\mu$. The corresponding results for $\Upsilon_{SDG}$ and $\Upsilon_{PDG}$ are presented in Fig.~\ref{figa5}((a)--(b)) and Fig.~\ref{figa5}((c)--(d)) (Appendix).  From Fig.~\ref{figwtheat12}, Fig.~\ref{figa5}(a), and Fig.~\ref{figa5}(c), it is evident that, whether in the entire system or within $\Upsilon_{SDG}$ and $\Upsilon_{PDG}$, for the same resource interval, the cooperation fraction corresponding to $\mu=0$ is the lowest. As $\mu$ increases, cooperation levels improve. Moreover, for a fixed $\mu$, the cooperation fraction in $\Upsilon_{SDG}$ is consistently the highest, while the cooperation fraction in $\Upsilon_{PDG}$ remains the lowest. The individual resource rank-plot in the entire system, $\Upsilon_{SDG}$, and $\Upsilon_{PDG}$ for different values of $\mu$ is shown in Fig.~\ref{figwtheat12}(b), Fig.~\ref{figa5}(b), and Fig.~\ref{figa5}(d), respectively. As $\mu$ increases, the disparity in individual resource distribution narrows, further confirming that $\mu$ enhances cooperation while simultaneously reducing resource inequality. To further promote social fairness and narrow the inequality gap, we introduce a minimum resource guarantee, represented by $\xi$. Under the combined effect of $\mu=0.5$ and $\xi=0.1$, the purple bar chart in Fig.~\ref{figwtheat12}(a), Fig.~\ref{figa5}(a), and Fig.~\ref{figa5}(c) illustrates the cooperation fraction for each resource-based division. Compared to the case of only $\mu=0.5$ (gray bar chart), the cooperation fraction increases significantly in higher resource intervals, both at the system level and within $\Upsilon_{SDG}$ and $\Upsilon_{PDG}$. Once the system stabilizes, the distribution of individual resources is represented by the purple line in Fig.~\ref{figwtheat12}(b), Fig.~\ref{figa5}(b), and Fig.~\ref{figa5}(d). Compared to other values of $\mu$, the number of individuals with higher resources decreases, and the threshold at which the purple line disappears is further elevated. This confirms that, whether in the entire system or within $\Upsilon_{SDG}$ and $\Upsilon_{PDG}$, the combined effect of $\mu$ and $\xi$ enhances cooperation and mitigates resource inequality among individuals.

\begin{figure}
    \centering
    \includegraphics[width=1.0\textwidth]{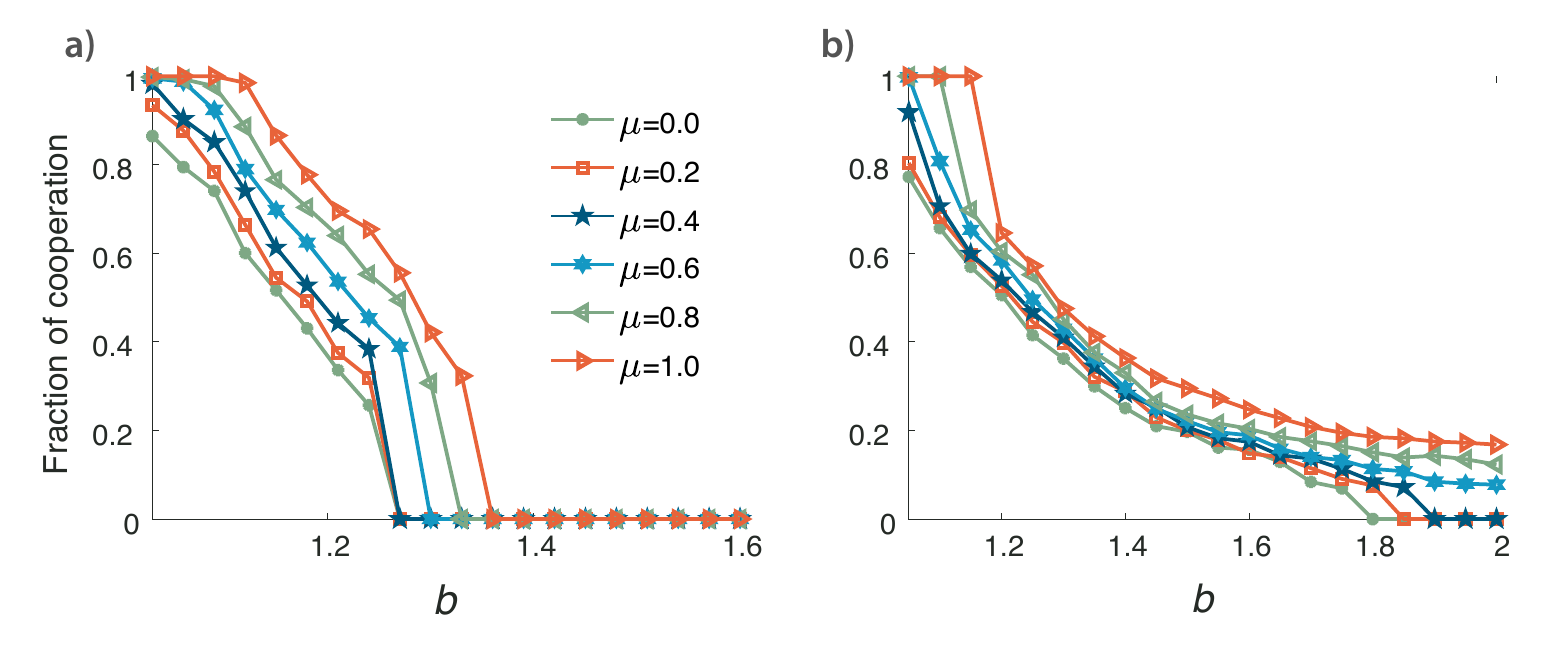}
    \caption{Robustness of social environment coupling parameter $\mu$ in promoting cooperation across different network structures. Cooperation fraction as a function of $b$ for varying values of $\mu$ in a small-world network (panel (a)) and a random graph network (panel (b)). The remaining control parameter is set to $\sigma=0.2$.}
    \label{figwtheat13}
\end{figure}

Finally, to test the robustness of our multi-game framework with adaptive control mechanism and limited resources, we analyze the impact of $\mu$ on cooperation levels in different network structures. Specifically, we examine the cooperation fraction as a function of $b$ in a small-world network (Fig.~\ref{figwtheat13}(a)) and a random graph network (Fig.~\ref{figwtheat13}(b)). As observed, in both network structures, increasing $\mu$ significantly enhances cooperation within the system and raises the threshold for the disappearance of cooperative behavior, which is consistent with previous findings. Furthermore, in Fig.~\ref{figwtheat13}(a), at approximately $b \approx 1.375$, cooperation within the small-world network increases regardless of the value of $\mu$. However, in the random graph network model (Fig.~\ref{figwtheat13}(b)), when $\mu\geq 0.6$, cooperative behavior persists across the entire system regardless of $b$. This suggests that, compared to the small-world network, $\mu$ is more effective in promoting cooperation in the random graph network model.

\section{Discussion}
\label{sec:discussion}

In this study, we proposed a multi-game model with limited resources that integrates an advantageous environmental mechanism, where individual fitness is dynamically redefined based on both self-payoff and the payoffs of social peers. This framework captures the co-evolution of strategies and resource allocation across two distinct game types: the Snowdrift Game (SDG) and the Prisoner’s Dilemma Game (PDG). By adopting an adaptive control approach, we examined system behavior across multiple resolution levels — from macroscopic trends to game-based subpopulations and microscopic configurations — and characterized both equilibrium states and transient dynamics.

Our results show that coupling individual behavior to an advantageous environment, alongside resource constraints, robustly promotes cooperation. Specifically, the interplay of the coupling parameter and the sucker’s payoff significantly raises the threshold at which cooperation persists, increases the prevalence of cooperative clusters, and mitigates early disadvantages faced by cooperators. And, the model reveals how richer resource endowments, shaped by strategic behavior and environmental context, reinforce cooperative tendencies and alter the distribution of cooperation across social groups. Moreover, from a system-level perspective, we observed that the interaction between resource availability and strategic dynamics yields unequal but structured resource distributions, with cooperators generally occupying higher-resource regimes. Introducing mechanisms such as a minimum resource guarantee further reduces inequality and enhances the persistence of cooperation. Finally, we confirmed that the advantageous environment mechanism remains robust across different network topologies, including small-world and random graphs.

Our findings underline the critical role of resource co-evolution and the individual's environment in shaping collective strategic behavior. While our model focuses on two canonical games, future work could explore its extension to more diverse interaction types, such as the Hawk-Dove Game or the Public Goods Game, and more intricate structures like multilayer or higher-order networks. These avenues could deepen our understanding of how environmental and structural factors jointly govern the emergence and sustainability of cooperation.

\section*{Acknowledgements}
This research is supported by the National Natural Science Foundation of China (Nos:72371052); China Scholarship Council under grant (Nos:202306060151).

\clearpage

\appendix
\renewcommand\thefigure{\Alph{section}\arabic{figure}} 
\setcounter{figure}{0}
\section{Supplementary figures}\label{secappendix}
Here, we provide some supplementary figures to support the findings in the main text.

%\begin{figure}[htbp]
\begin{figure}[H]
    \centering
    \begin{subfigure}{0.9\textwidth}
        \centering
        \includegraphics[width=\linewidth]{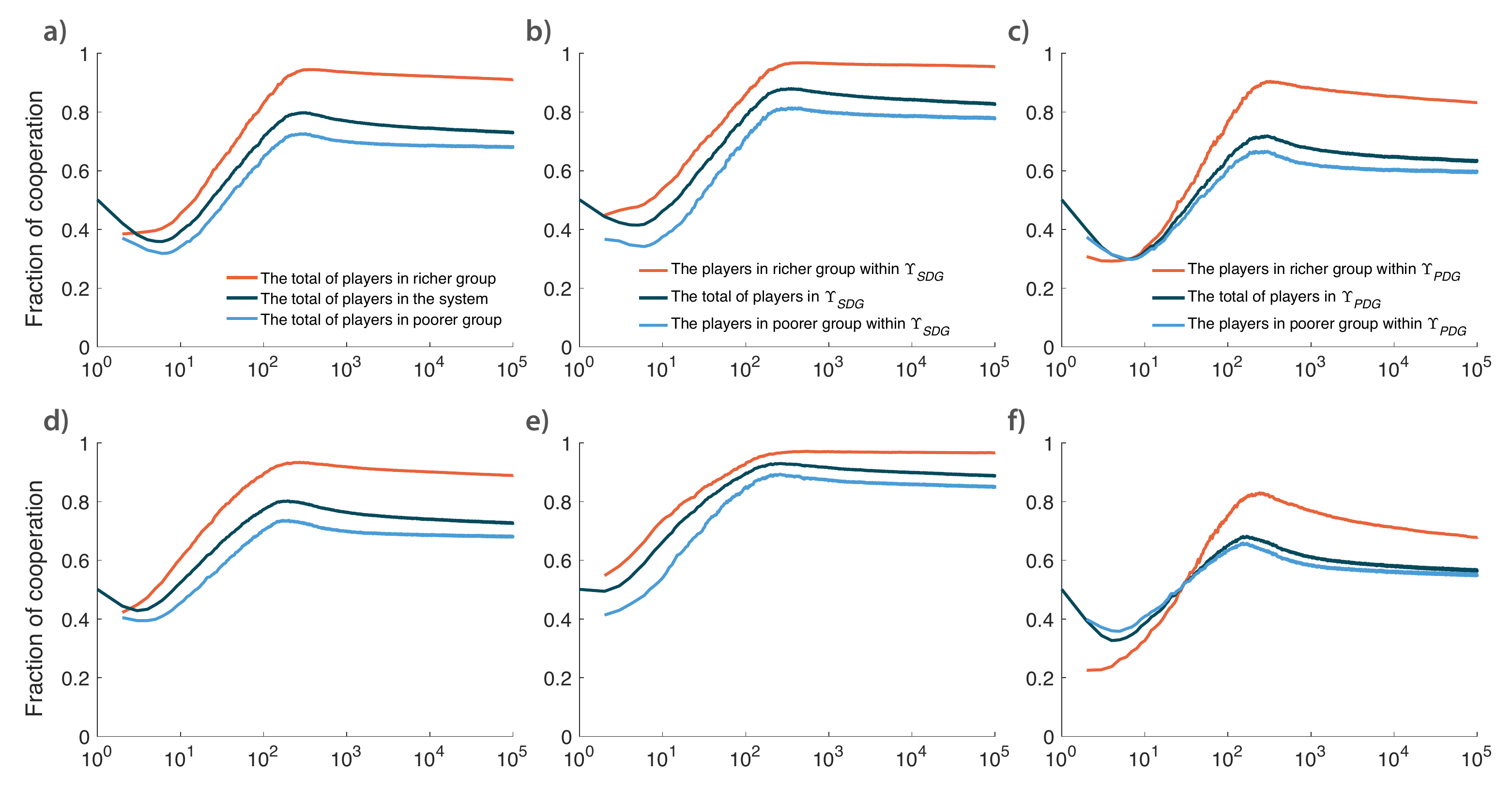}
        % \caption{(a)}
    \end{subfigure}
    \vspace{0cm}
    \begin{subfigure}{0.9\textwidth}
        \centering
        \includegraphics[width=\linewidth]{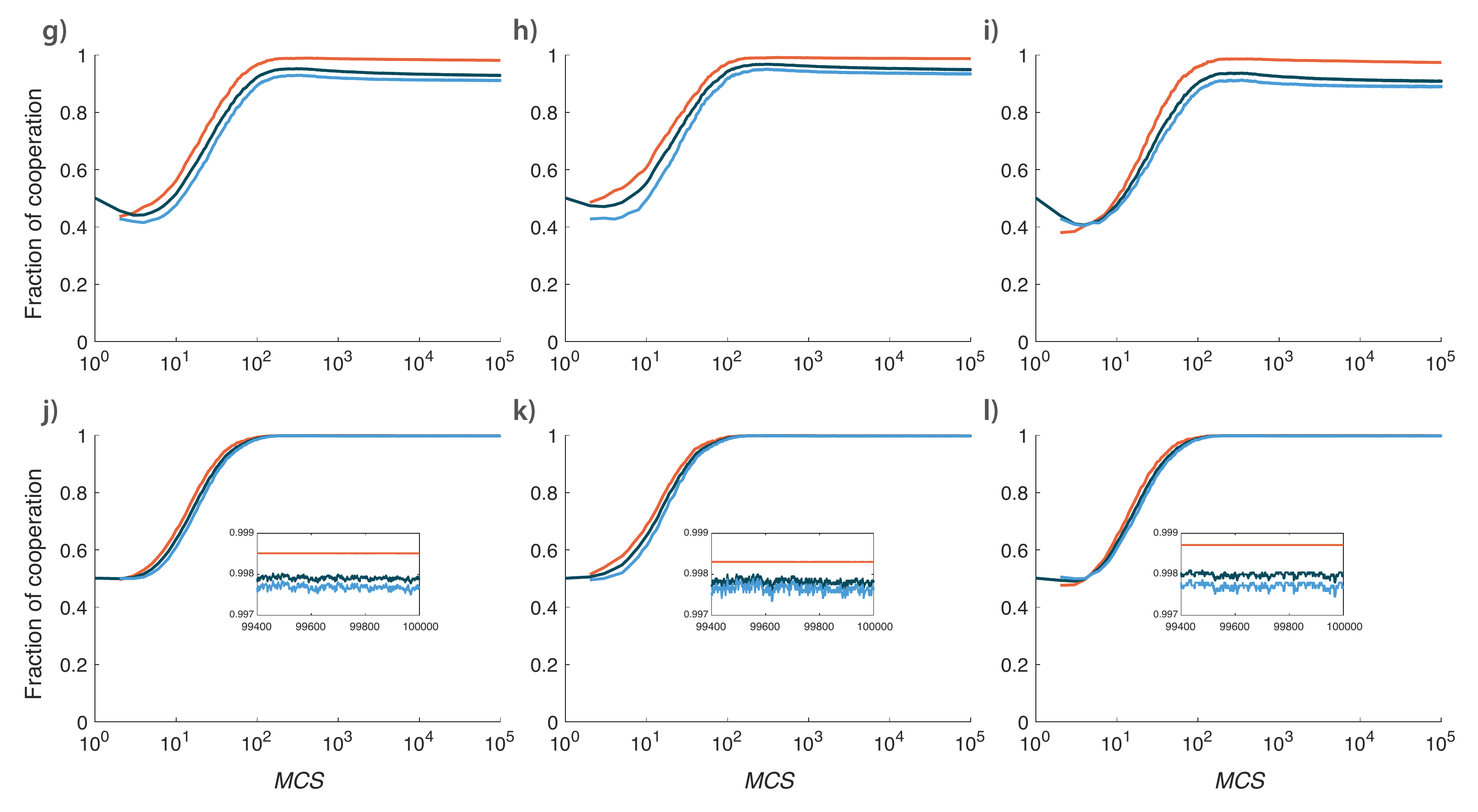}
    \end{subfigure}
    \caption{Time evolution in Monte Carlo Steps (MCS) of the cooperation fraction in the richer and poorer groups, classified based on individuals’ resource quantity. The richer group consists of individuals with resources greater than $1$ unit, while the poorer group consists of individuals with resources less than $1$ unit. Left column (panels (a)--(d)): Cooperation fraction in the richer and poorer groups for the entire system. Central column (panels (e)--(h)) and right column (panels (i)--(l)) presents the corresponding evolution within within $\Upsilon_{SDG}$ and $\Upsilon_{PDG}$. The top-to-bottom arrangement corresponds to $\mu=0$, $0.2$, $0.8$, $1.0$, respectively. The remaining control parameters are set to $\sigma=0.2$ and $b=1.10$.}
    \label{figa1}
\end{figure}

%\begin{figure}[htbp]
\begin{figure}[H]
    \centering
    \begin{subfigure}{0.9\textwidth}
        \centering
        \includegraphics[width=\linewidth]{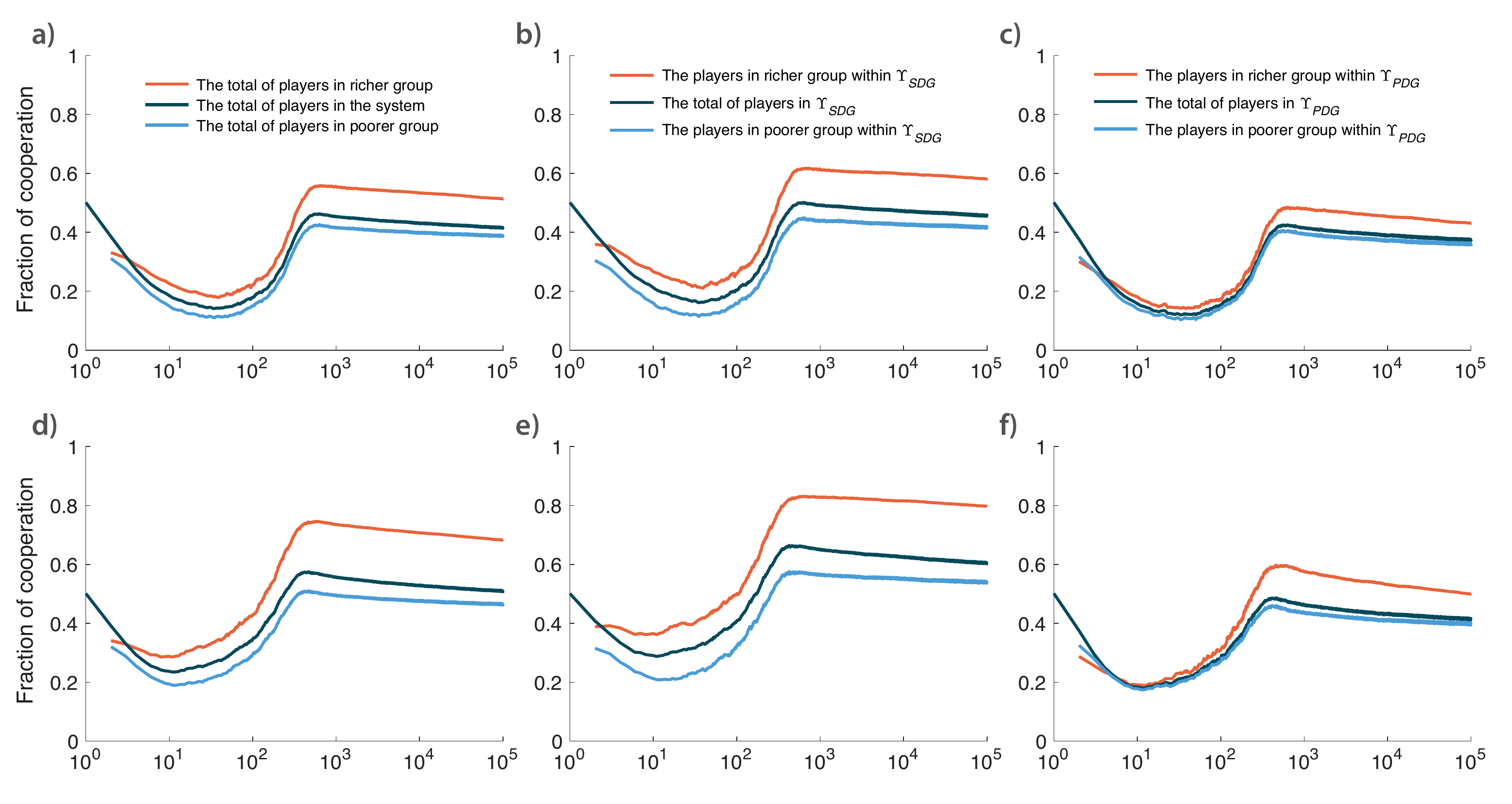}
    \end{subfigure}
    \vspace{0cm}
    \begin{subfigure}{0.9\textwidth}
        \centering
        \includegraphics[width=\linewidth]{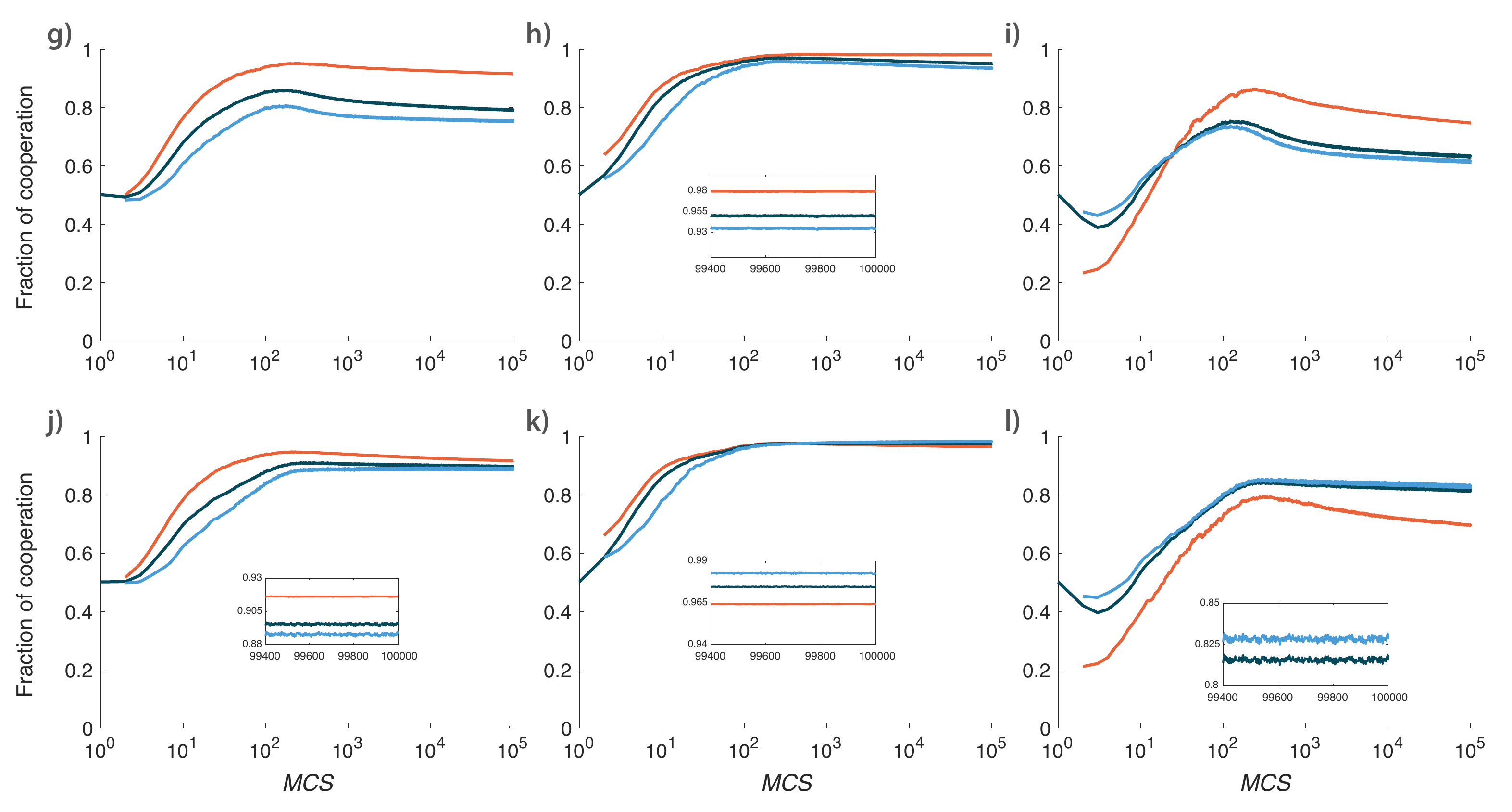}
    \end{subfigure}
    \caption{Time evolution in Monte Carlo Steps (MCS) of the cooperation ratio in the richer and poorer groups, classified based on individuals’ resource quantity. The richer group consists of individuals with resources greater than $1$ unit, while the poorer group consists of individuals with resources less than $1$ unit. Left column (panels (a)--(d)): Cooperation ratio in the richer and poorer groups for the entire system. Central column (panels (e)--(h)) and right column (panels (i)--(l)) presents the corresponding
    evolution within within $\Upsilon_{SDG}$ and $\Upsilon_{PDG}$. The top-to-bottom arrangement corresponds to $\sigma=0.05$, $0.1$, $0.8$, $1.0$, respectively. The remaining control parameters are set to $\mu=0.2$ and $b=1.10$.}
    \label{figa2}
\end{figure}

%\begin{figure}[htbp]
\begin{figure}[H]
    \centering
    \begin{subfigure}{0.9\textwidth}
        \centering
        \includegraphics[width=\linewidth]{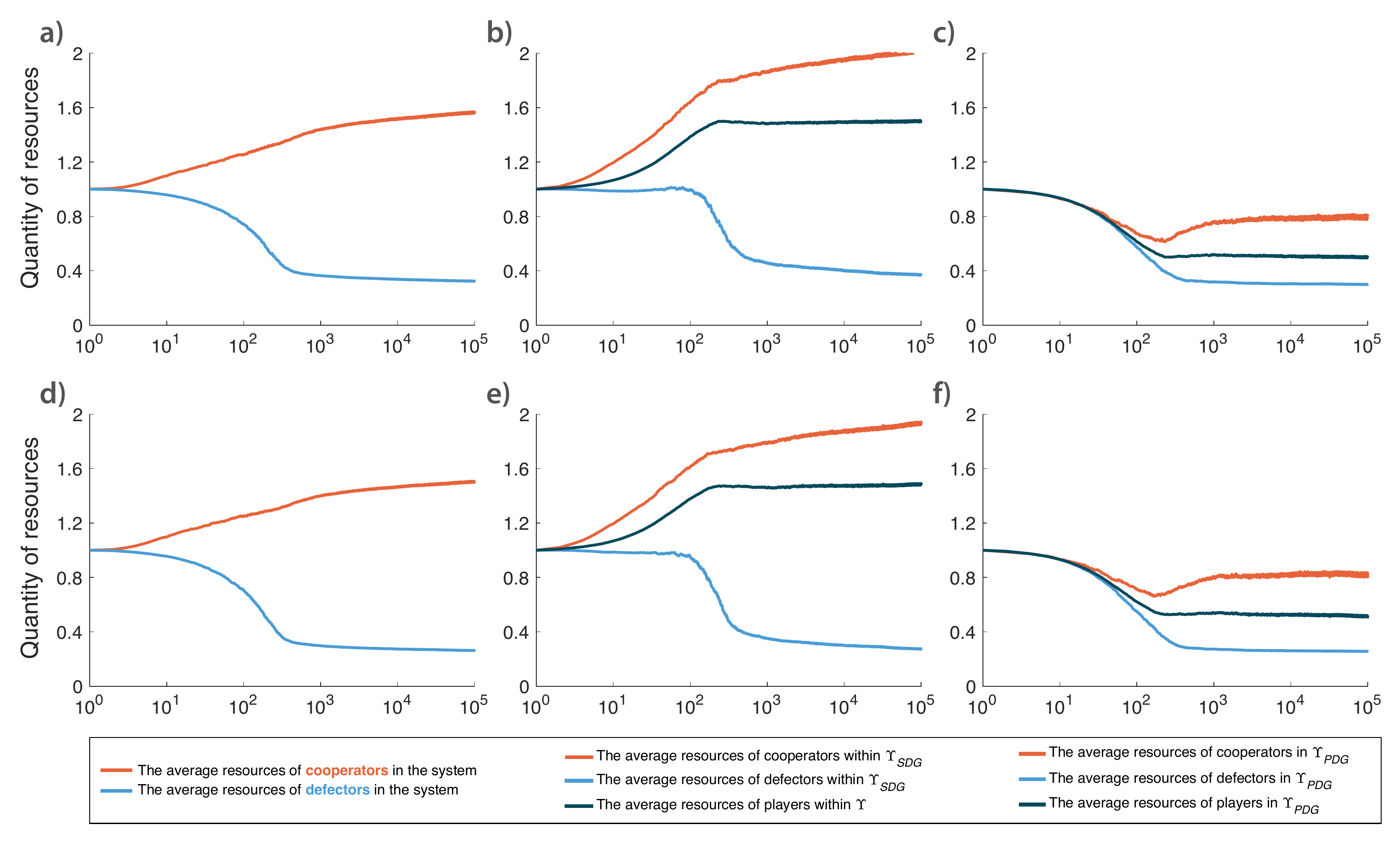}
    \end{subfigure}
    \vspace{0cm} 
    \begin{subfigure}{0.9\textwidth}
        \centering
        \includegraphics[width=\linewidth]{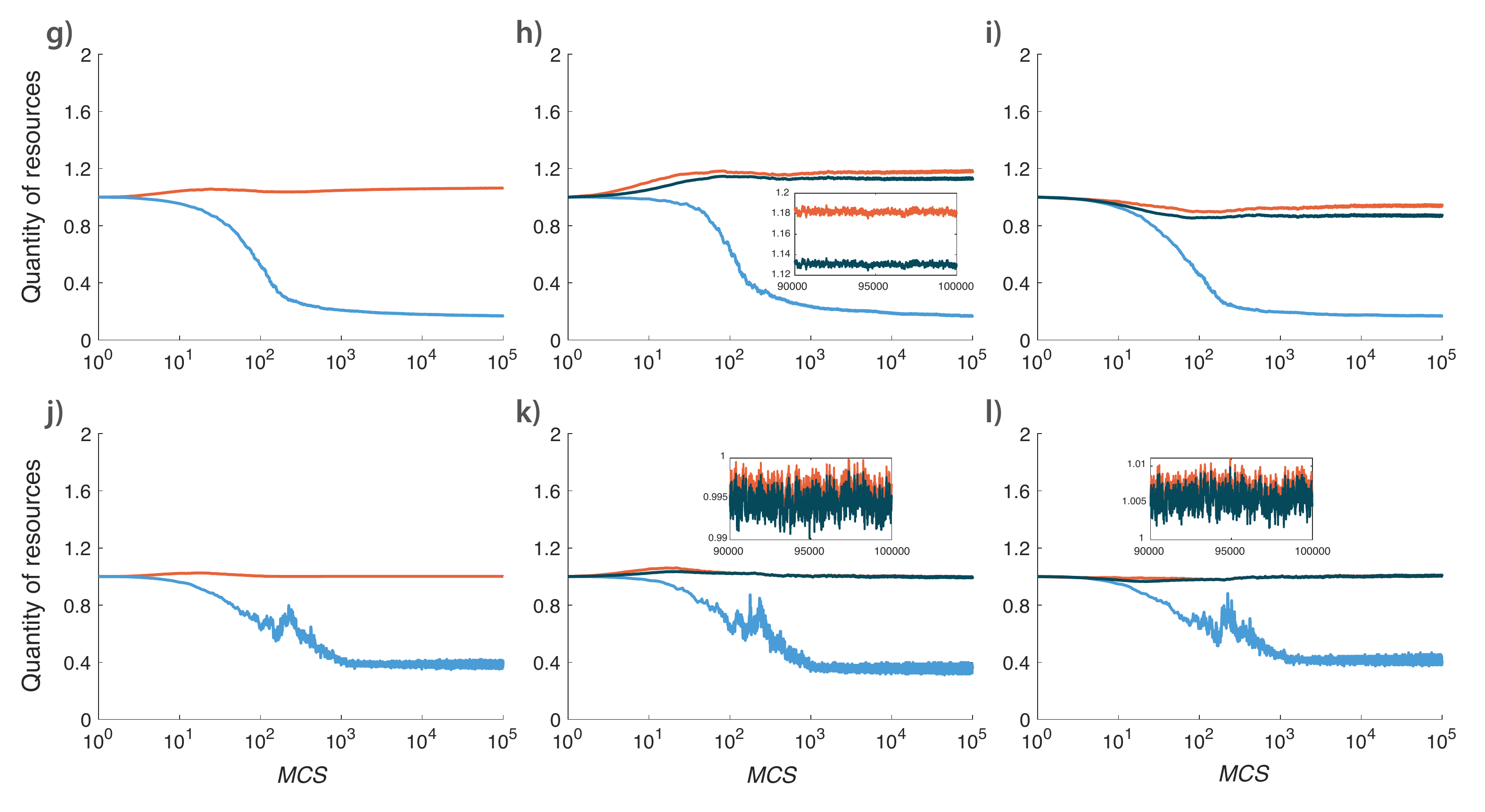}
    \end{subfigure}
    \caption{Time evolution in Monte Carlo Steps (MCS) of the average resources of cooperators and defectors in the entire system, as well as within $\Upsilon_{SDG}$ and $\Upsilon_{PDG}$. Left column (panels (a), (d), (g), (j)): Evolution of the average resources of cooperators and defectors in the entire system. Central column (panels (b), (e), (h), (k)) and  right column (panels (c), (f), (i), (l)) presents the corresponding evolution within $\Upsilon_{SDG}$ and $\Upsilon_{PDG}$. The top-to-bottom arrangement corresponds to $\mu=0$, $0.2$, $0.8$, $1.0$, respectively. The remaining control parameters are set to $\sigma=0.2$ and $b=1.10$.}
    \label{figa3}
\end{figure}

\begin{figure}[H]
    \centering
    \begin{subfigure}{0.9\textwidth}
        \centering
        \includegraphics[width=\linewidth]{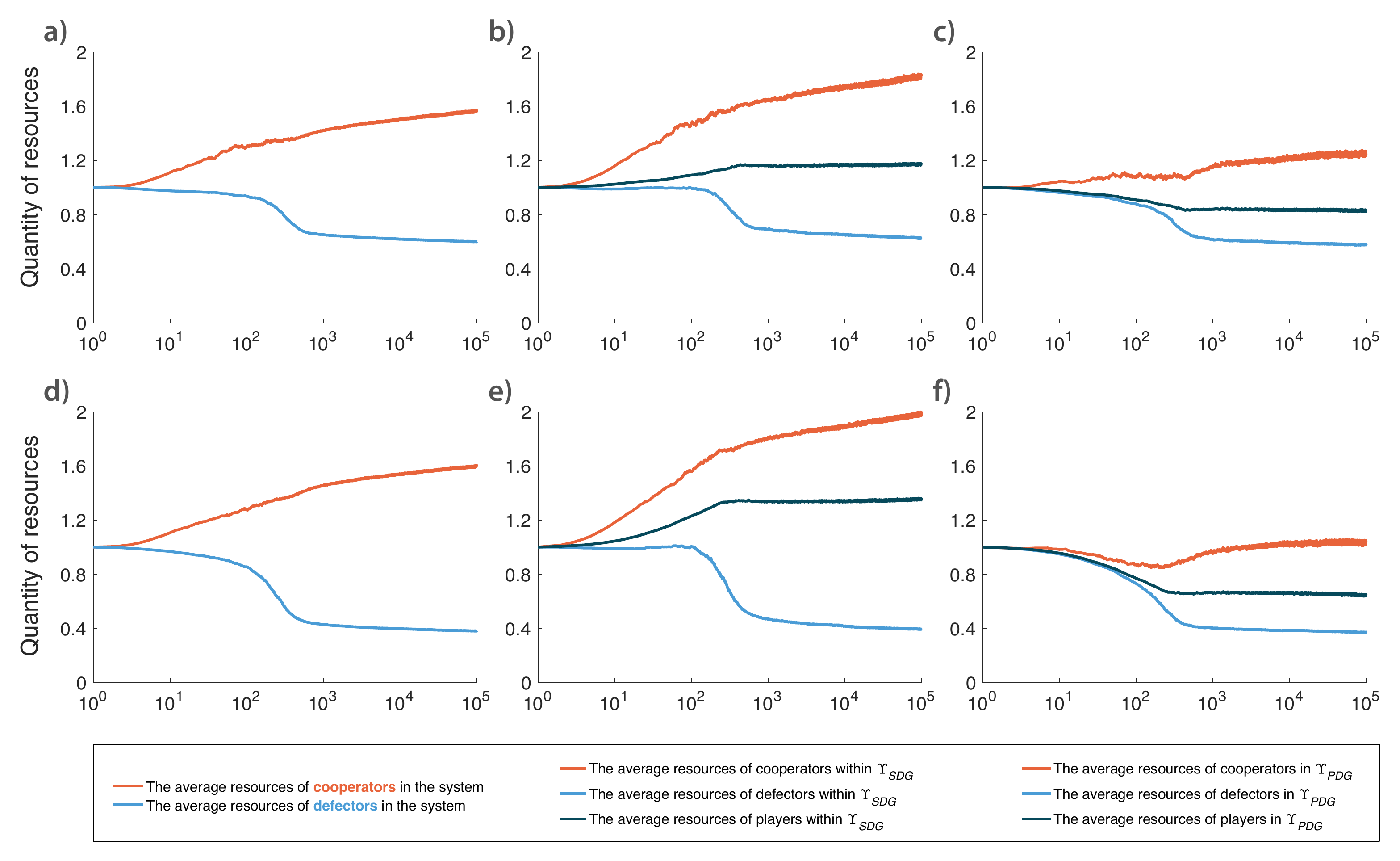}
    \end{subfigure}
    \vspace{0cm} 
    \begin{subfigure}{0.9\textwidth}
        \centering
        \includegraphics[width=\linewidth]{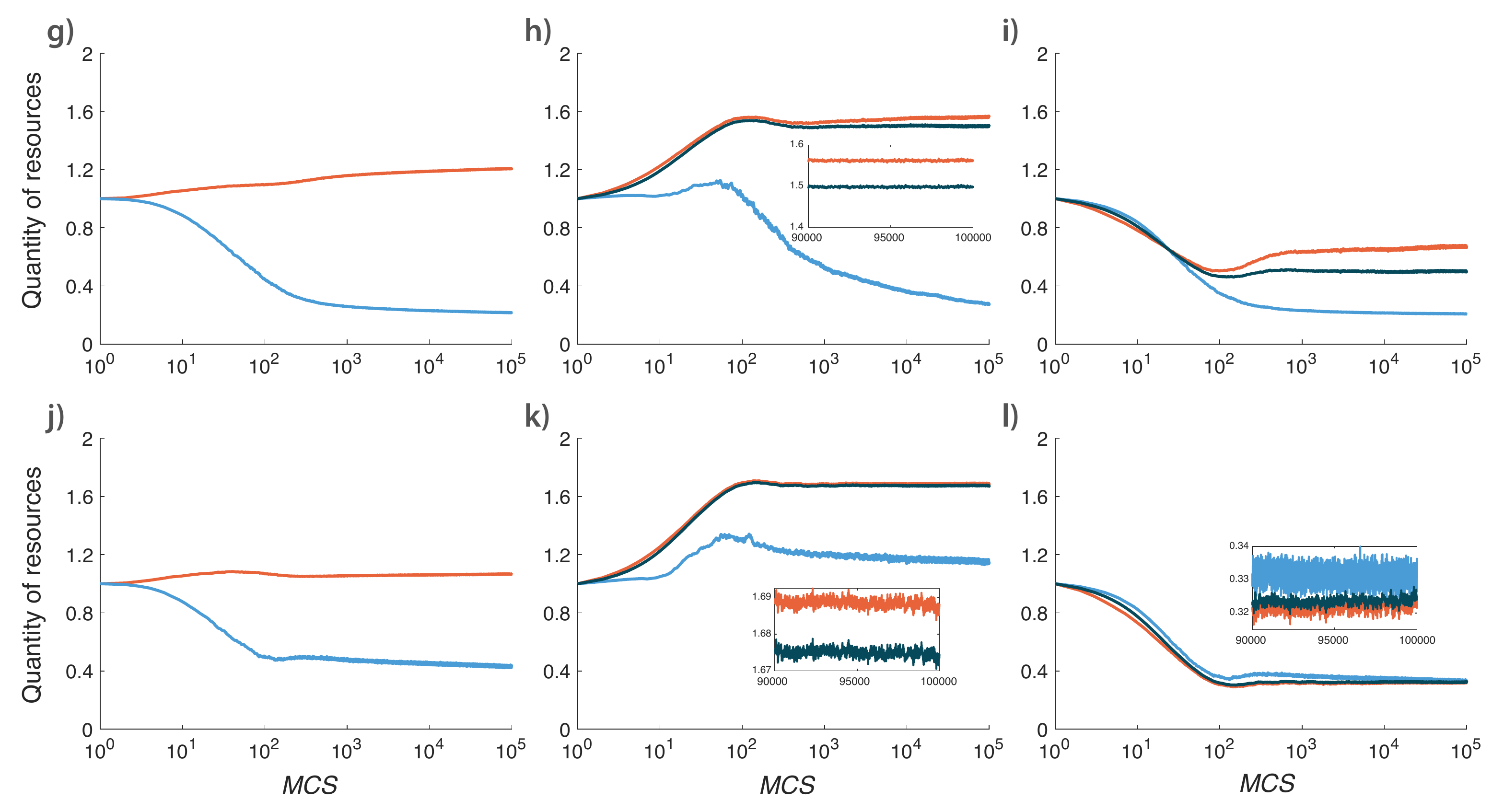}
    \end{subfigure}
    \caption{Time evolution in Monte Carlo Steps (MCS) of the average resources of cooperators and defectors in the entire system, as well as within $\Upsilon_{SDG}$ and $\Upsilon_{PDG}$. Left column (panels (a), (d), (g), (j)): Evolution of the average resources of cooperators and defectors in the entire system. Central column (panels (b), (e), (h), (k)) and  right column (panels (c), (f), (i), (l)) presents the corresponding evolution within $\Upsilon_{SDG}$ and $\Upsilon_{PDG}$. The top-to-bottom arrangement corresponds to $\sigma=0.05$, $0.1$, $0.8$, $1.0$, respectively. The remaining control parameters are set to $\mu=0.2$ and $b=1.10$.}
    \label{figa4}
\end{figure}

\begin{figure}[h]
    \centering
    \makebox[\textwidth][l]{\hspace{-0.08\textwidth} % [l] aligns left within the textwidth
        \begin{adjustbox}{max width=\textwidth}
            \begin{minipage}{1.0\textwidth}
                % Top row
                \begin{subfigure}{0.67\textwidth}
                    \centering
                    \includegraphics[height=5.2cm]{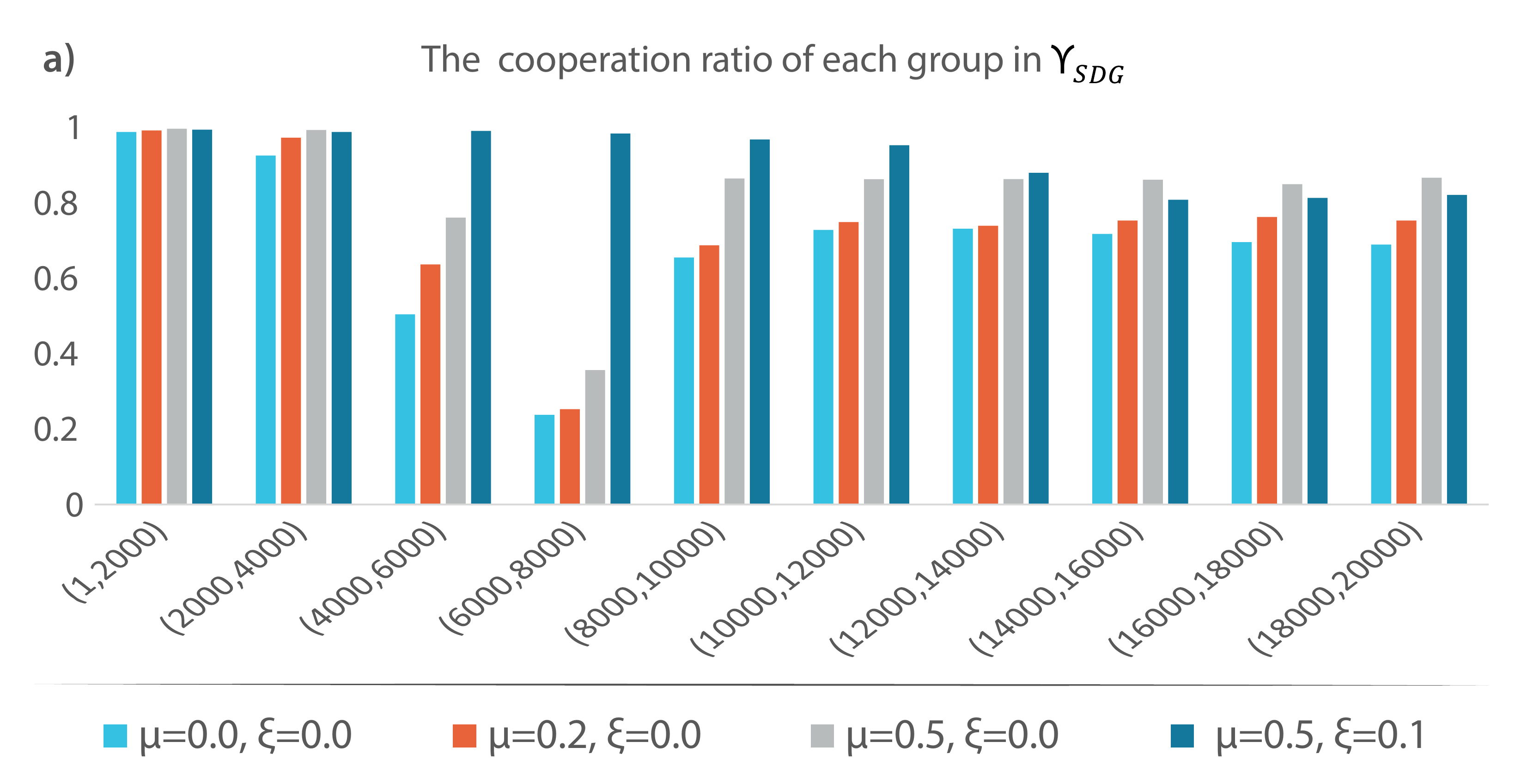}
                \end{subfigure}
                \hfill
                \begin{subfigure}{0.3\textwidth}
                    \centering
                    \includegraphics[height=5.2cm]{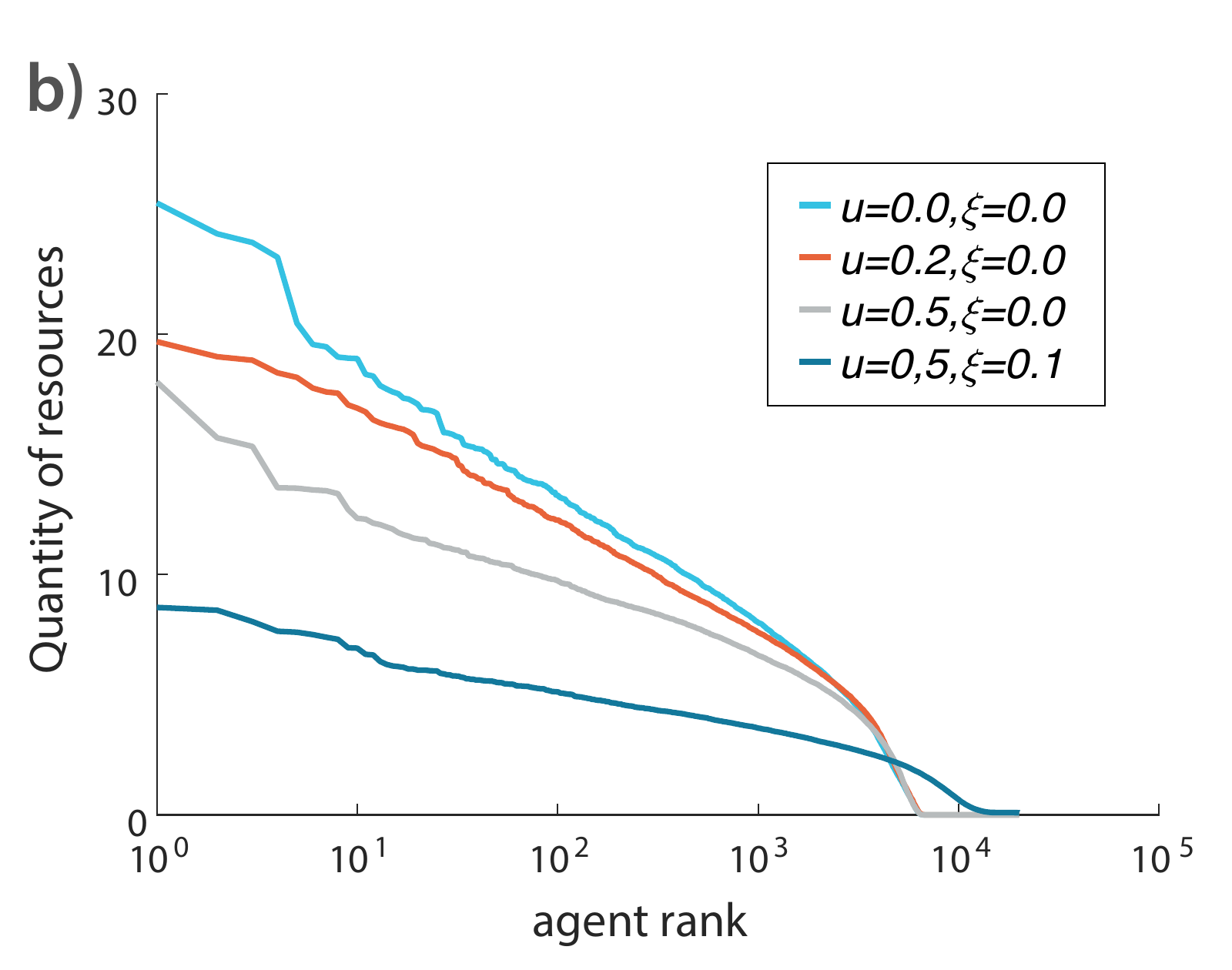}
                \end{subfigure}
                %\vspace{0.5em}
                % Bottom row
                \begin{subfigure}{0.67\textwidth}
                    \centering
                    \includegraphics[height=5.2cm]{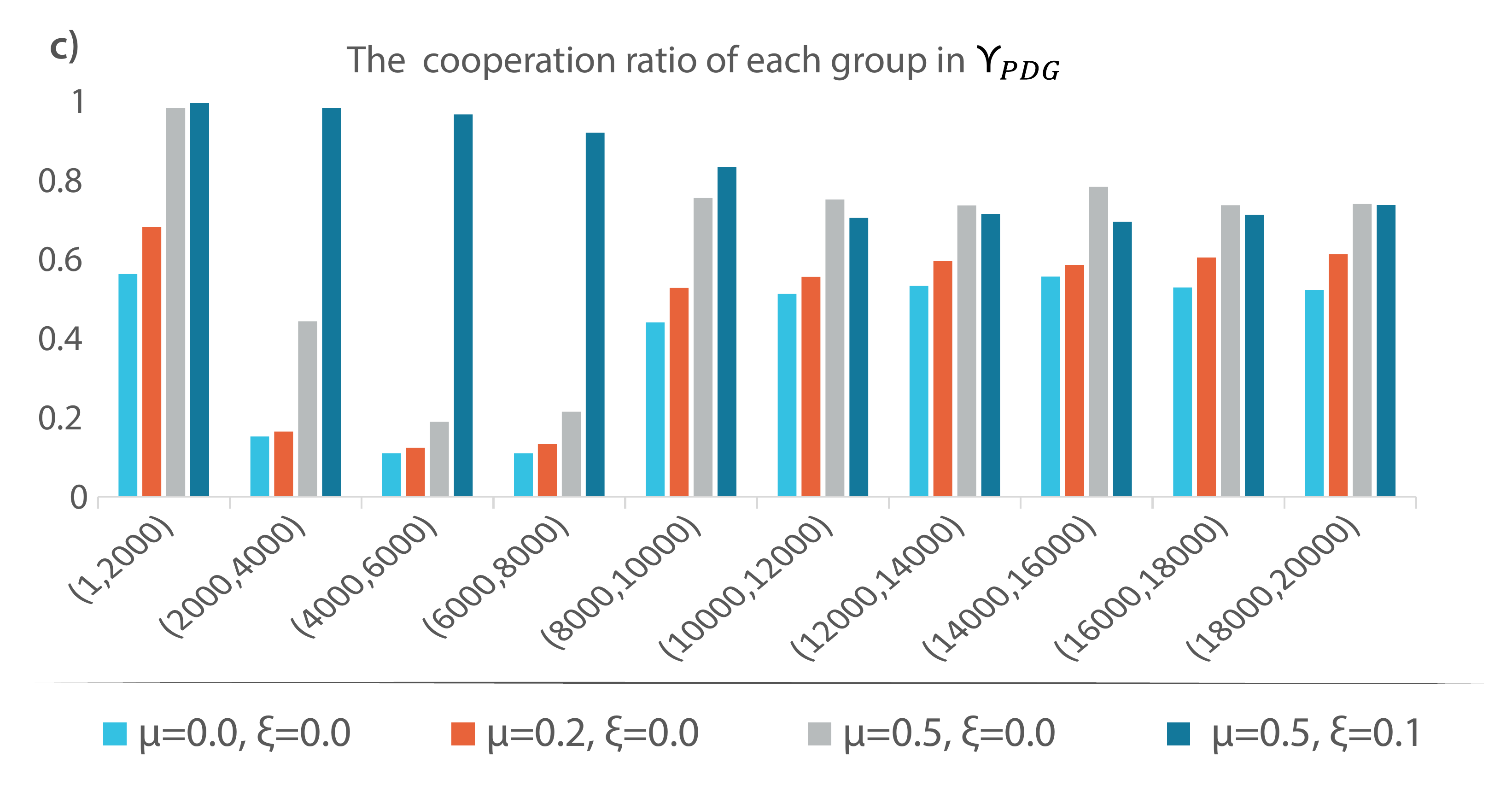}
                \end{subfigure}
                \hfill
                \begin{subfigure}{0.3\textwidth}
                    \centering
                    \includegraphics[height=5.2cm]{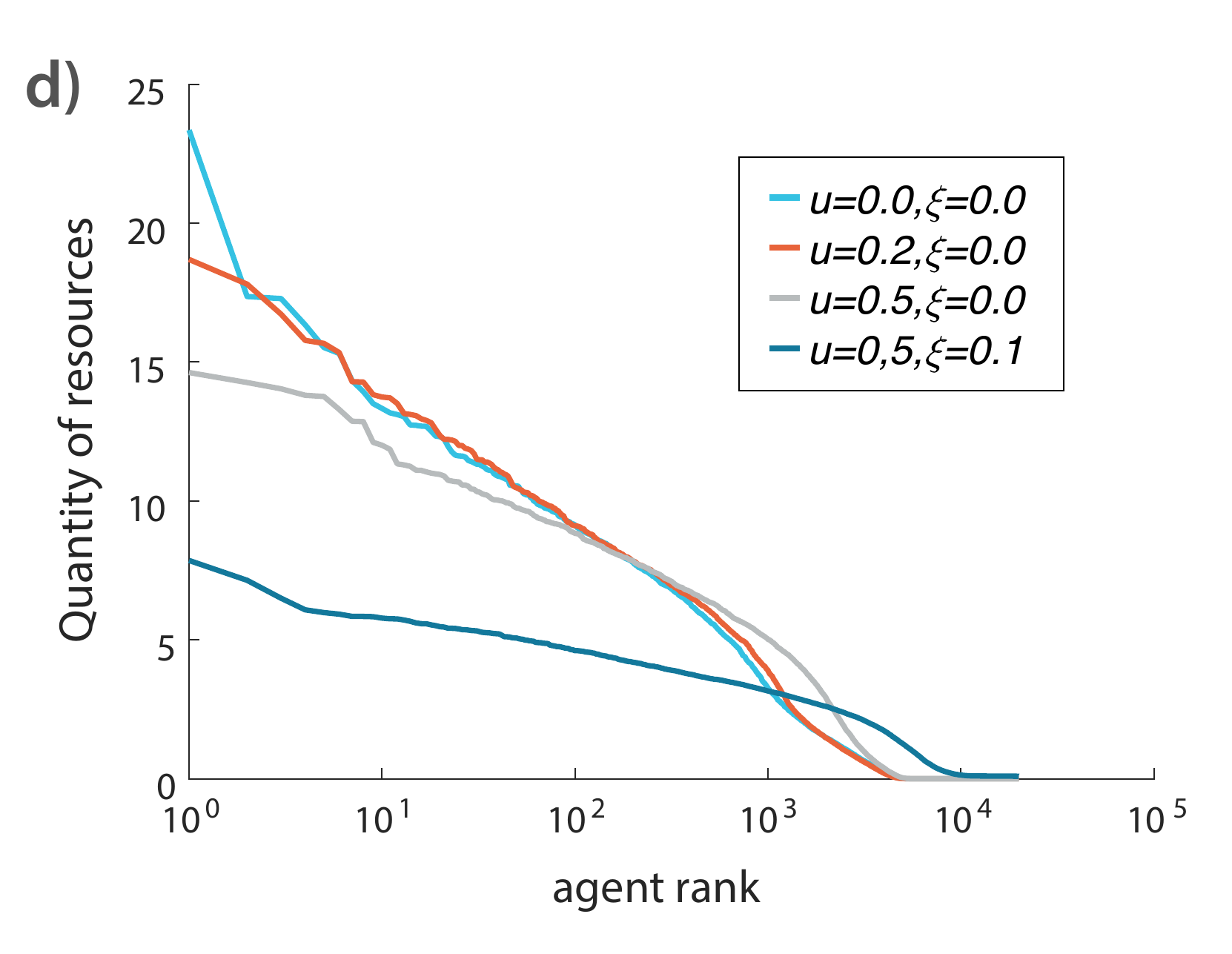}
                \end{subfigure}
            \end{minipage}
        \end{adjustbox}
    }
    \caption{Cooperation fraction and resource rank-plot across different resource-based groups when the system reaches the equilibrium state. Players in the entire system, players in $\Upsilon_{SDG}$, and in $\Upsilon_{PDG}$ are evenly divided into ten groups based on their resource quantity. Left column (panels (a) and (c)): Cooperation fraction of each resource-based group for the entire system and within $\Upsilon_{SDG}$ and $\Upsilon_{PDG}$. Right column (panels (b) and (d)): Resource rank-plot of individuals in the entire system, $\Upsilon_{SDG}$, and $\Upsilon_{PDG}$. Panel (a) and panel (c) show results for different values of $\mu$, where $\mu=0$, $\mu=0.2$, $\mu=0.5$, and $\mu=0.5$ with $\xi=0.1$, where $\xi$ represents the resource guarantee parameter. The corresponding resource rank-plots for the system, $\Upsilon_{SDG}$, and $\Upsilon_{PDG}$ are presented in panel (b) and panel (d). The remaining control parameters are set to $\sigma=0.2$ and $b=1.10$.}
    \label{figa5}
\end{figure}

\clearpage

%\nocite{*}

\bibliography{references}

\end{document}